\pdfoutput=1

\documentclass[11pt]{article}

\usepackage[preprint]{acl}

\usepackage{times}
\usepackage{latexsym}

\usepackage[T1]{fontenc}

\usepackage[utf8]{inputenc}

\usepackage{microtype}

\usepackage{inconsolata}

\usepackage{graphicx}

\title{Roles of MLLMs in Visually Rich Document Retrieval for RAG: A Survey}

\author{
    Xiantao Zhang \\
    Beihang University \\
    \texttt{zhangxiantao@buaa.edu.cn}
}

\usepackage{amsmath}
\usepackage{amsfonts}
\usepackage{booktabs}
\usepackage{multirow}
\usepackage{enumitem}
\usepackage[table]{xcolor}
\definecolor{LightCyan}{RGB}{224,255,255}

\begin{document}
\maketitle

\begin{abstract}

Visually rich documents (VRDs) challenge retrieval-augmented generation (RAG) with layout-dependent semantics, brittle OCR, and evidence spread across complex figures and structured tables.
This survey examines how Multimodal Large Language Models (MLLMs) are being used to make VRD retrieval practical for RAG.
We organize the literature into three roles: \emph{Modality-Unifying Captioners}, \emph{Multimodal Embedders}, and \emph{End-to-End Representers}.
We compare these roles along retrieval granularity, information fidelity, latency and index size, and compatibility with reranking and grounding.
We also outline key trade-offs and offer some practical guidance on when to favor each role.
Finally, we identify promising directions for future research, including adaptive retrieval units, model size reduction, and the development of evaluation methods.

\end{abstract}

\section{Introduction}
\label{sec:introduction}

Visually rich documents (VRDs), such as PDFs, scanned pages, slide decks, reports, forms, and infographics, encode meaning through the interplay of text, layout, figures, and graphics.
As retrieval-augmented generation (RAG) \cite{lewis2020rag} becomes a default pattern for grounding large language models (LLMs) \cite{guu2020REALM, borgeaud2022RETRO, izacard2023ATLAS, nakano2022webgpt}, many real-world deployments are moving beyond plain text to these document types \cite{ma2024dse, faysse2025colpali, yu2025visrag, suri2025visdom, Tanaka2025VDocRAG}.
This shift strains classical text-only RAG pipelines, motivating the broader development of multimodal RAG (MM-RAG) systems designed to retrieve and reason over varied data types, including images and tables \cite{chen2022murag, yasunaga2023retrievalaugmentedmultimodallanguagemodeling}.

However, VRDs represent a uniquely difficult case for MM-RAG.
Unlike retrieving standalone images or text, VRD retrieval must contend with meaning derived from the \emph{fusion} of layout, embedded text, and graphics.
Consequently, traditional preprocessing steps like optical character recognition (OCR) and layout parsing remain brittle and lossy, fine-grained visual cues vanish in textual proxies, and evidence may span multiple pages or views.
Recent surveys in document understanding \cite{ding2025deeplearningbasedvisually} echo this shift, underscoring both the opportunity and difficulty of learning over text–layout–vision jointly.

At the same time, a new generation of methods argues for \emph{seeing} pages directly.
Document Screenshot Embedding (DSE) \cite{ma2024dse} treats a page screenshot as the unit of indexing, avoiding preprocessing choices that introduce error and latency. Its premise is pragmatic: keep all information available at retrieval time.
Likewise, ColPali \cite{faysse2025colpali} fuses Vision–Language Models (VLMs) with late-interaction matching, and results show that learning directly over page images can simplify pipelines and improve effectiveness.
Beyond retrieval alone, VisRAG \cite{yu2025visrag} integrates vision-based retrieval with generation, adopting the page-as-image abstraction end-to-end to mitigate conversion loss during both retrieval and answer synthesis.

VRD-centric evaluation has also matured. Beyond classic DocVQA \cite{mathew2021docvqa}, InfographicVQA \cite{Mathew2022InfographicVQA}, and SlideVQA \cite{Tanaka2023SlideVQA}, newer resources now increasingly stress chart reasoning and multi-slide evidence aggregation reflecting practical needs like finding a single number inside a plot or tracing an argument across a deck \cite{tamber2025benchmarkingllmfaithfulnessrag, yang2025benchmarkingmultimodalragchartbased, liu2025benchmarkingretrievalaugmentedgenerationmultimodal, chen2025VisRBench, peng2025UniDocBench}.
These datasets collectively highlight why retrieval must respect both layout and visual semantics, not only text.

\paragraph{Scope and goal}
This survey focuses specifically on visually rich document retrieval for RAG.
We analyze how Multimodal Large Language Models (MLLMs) are used to index and retrieve pages, page regions, tables, figures, and slide content for RAG over documents.
Our goal is to distill design patterns, compare empirical trends, and surface trade-offs that matter for building reliable, cost-aware systems.

\paragraph{Contributions}
This survey makes the following contributions:
\begin{enumerate}
    \item \textbf{Role-based taxonomy of VRD–RAG.} We organize how MLLMs enter the pipeline into three roles tailored to documents.
    \item \textbf{Comparative analysis of key trade-offs.} We contrast these roles in terms of retrieval unit, robustness to OCR and layout errors, latency and indexing cost, and compatibility with reranking and grounding, summarizing evidence from recent VRD-focused work.
    \item \textbf{Practical takeaways and open challenges.} We discuss when to favor caption-first vs. image-first retrieval, how to balance page-level recall with element-level precision, how to budget compute and storage for multimodal indices, and where evaluation lags behind given the current benchmarks.
\end{enumerate}

\paragraph{Organization}
\S\ref{sec:related_work} reviews background on RAG, multimodal retrieval, and MLLMs. \S\ref{sec:roles_of_mllm_in_vrd_rag} develops the three-role framework and contrasts representative approaches. \S\ref{sec:trade_offs} examines trade-offs and open challenges. \S\ref{sec:conclusion} concludes with takeaways and future directions.

\begin{figure*}[ht]
  \centering
  \includegraphics[width=\linewidth]{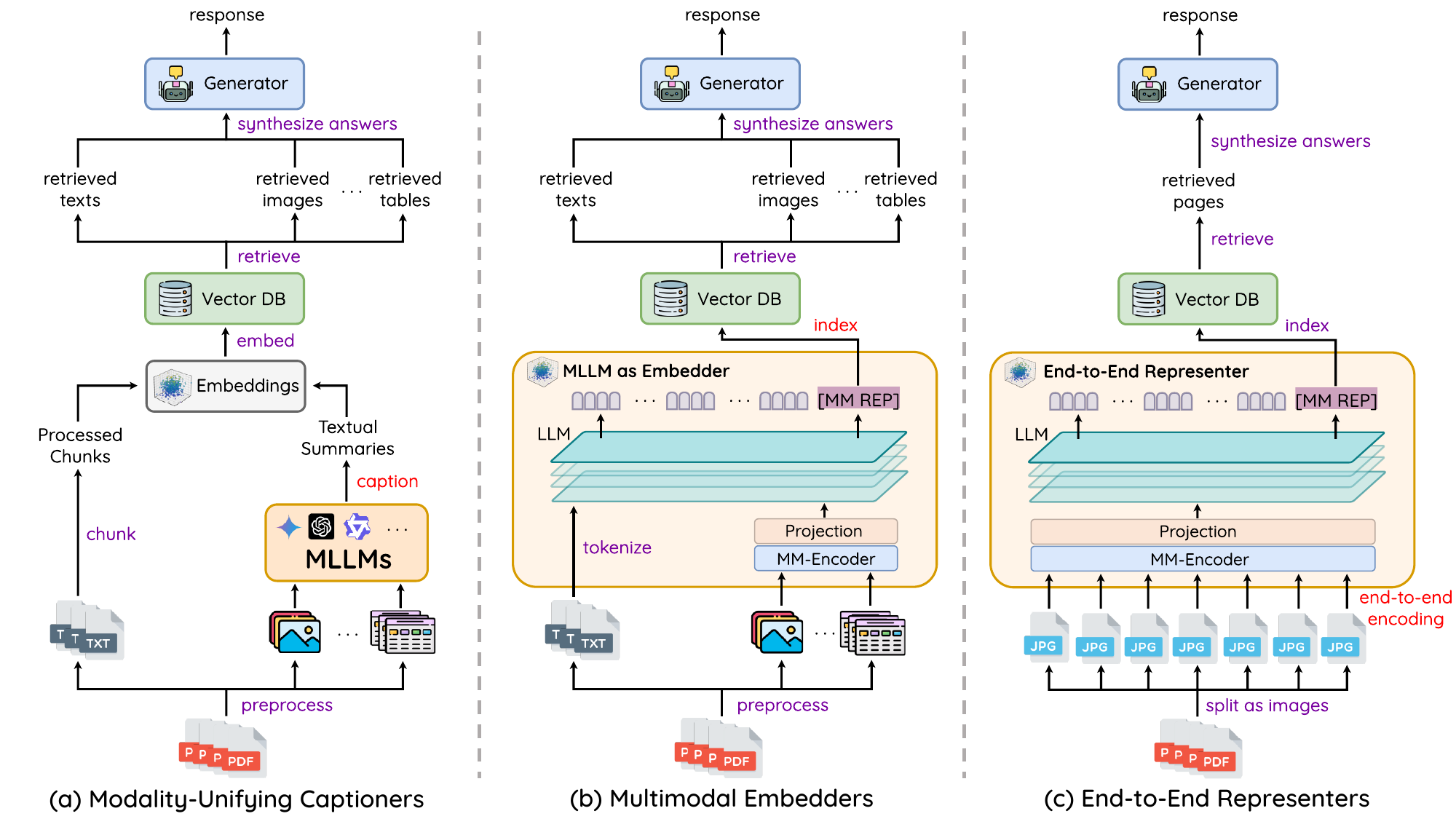}
  \caption{Overview of how MLLMs enter VRD retrieval for RAG across three roles.
  \textbf{Left:} \emph{Modality-Unifying Captioners} (\S\ref{subsec:mllm-as-captioner});
  \textbf{Middle:} \emph{Multimodal Embedders} (\S\ref{subsec:mllm-as-mmemb});
  \textbf{Right:} \emph{End-to-End Representers} (\S\ref{subsec:mllm-as-e2erepr}).
  Each panel sketches the pipeline from document intake to retrieval and answer synthesis, highlighting typical retrieval units and index types.}
  \label{fig:main-fig}
\end{figure*}

\section{Related Work}
\label{sec:related_work}

\subsection{Retrieval-Augmented Generation}
\label{subsec:rag}

RAG combines a retriever and a generator to bridge retrieval-based and generative models.
This hybrid approach dynamically retrieves documents to condition generation, enhancing factual accuracy and access to knowledge beyond training data \cite{shuster2021ragreduceshallucination, gao2024ragforllms_survey, lewis2020rag, asai2024selfrag, shi2024replug, izacard2023ATLAS, borgeaud2022RETRO, li2024enhancingllmfactualaccuracy}.
Recent research expanded RAG to open-domain QA \cite{guu2020REALM, mao2021generationaugmentedretrievalopendomainquestion, siriwardhana2023improvingdomainadaptionofrag}, dialogue systems \cite{thulke2021efficientretrievalaugmentedgeneration, komeili2022internetaugmenteddialoguegeneration, li2022knowledgegroundeddialoguegeneration}, and multimodal tasks \cite{chen2022murag, yasunaga2023retrievalaugmentedmultimodallanguagemodeling, Hu2023REVEAL, luo2024videorag, ren2025videorag, jeong2025videorag}, highlighting its potential for integrating diverse knowledge into NLP pipelines.

\subsection{Multimodal Retrieval}
\label{subsec:multmodal_retrieval}

Recent studies have demonstrated that multimodal retrieval and RAG significantly enhance LLMs by integrating diverse data modalities, such as text, images, and audio.
The seminal work of MuRAG \cite{chen2022murag} inaugurated the era of end-to-end multimodal retrieval-augmented transformers, a pioneering innovation that has since been shown to enhance performance in a range of tasks, including question answering, by leveraging external multimodal memory.
In a similar manner, RA-CM3 \cite{yasunaga2023retrievalaugmentedmultimodallanguagemodeling} was the first to demonstrate the capabilities of joint retrieval and text and image generation, achieving superior performance compared to models such as DALL-E \cite{ramesh2021dalle}, while being more efficient.
\citet{wei2024uniir} proposed UniIR, a universal multimodal retrieval model designed to handle a wide range of tasks.
Subsequent advancements include GENIUS \cite{kim2025geniusgenerativeframeworkuniversal}, a universal generative framework for multimodal search, and UMaT \cite{bi2025describedwordssimpleunified}, which unifies video and audio data via textual representations for long-form question-answering.
A comprehensive survey by \citet{zhao2023retrievingmultimodalinformationaugmented} further systematizes these approaches, highlighting improvements in factuality, robustness, and cross-modal reasoning.
Collectively, these works emphasize the transformative potential of multimodal RAG in scaling LLM capabilities across domains.

\subsection{Multimodal Large Language Models}
\label{subsec:mllm}

MLLMs have emerged as a transformative advancement in the field of artificial intelligence, extending the capabilities of LLMs by integrating multiple data modalities, such as text, images, audio, and videos.
LLaVA \cite{liu2023llava, liu2024llavanext} has been at the forefront of visual instruction tuning, achieving this through the alignment of a vision encoder with a language model via a cross-modal connector.
Subsequent developments like the Qwen-VL Series \cite{bai2023qwenvl, wang2024qwen2vl, bai2025qwen2_5vl} and the InternVL Series \cite{chen2024internvl, chen2024internvl1_5, chen2025internvl2_5, zhu2025internvl3, wang2025internvl3_5} have demonstrated significant progress in multimodal understanding and reasoning, including specialized alignment techniques for complex domains like mathematical reasoning \cite{zhuang2025mathpuma}.

\subsection{Related Surveys}
\label{subsec:related_surveys}

A growing body of surveys maps the RAG landscape from text-only pipelines to fully multimodal systems.
Early overviews on RAG \cite{gao2024ragforllms_survey, fan2024asurveyonragmeetingllms} consolidate architectures, training strategies, and evaluation, motivating retrieval as a remedy for hallucinations and stale knowledge. 
\citet{xu2025surveymodelarchitecturesinformation} survey the evolution of model
architectures in information retrieval (IR).
For multimodality, \citet{zhao2023retrievingmultimodalinformationaugmented} provide one of the first broad treatments across images, tables, and audio.
More recent efforts expand the scope and depth: \citet{abootorabi2025askinanymodality} organize the full multimodal RAG pipeline together with datasets and training strategies; \citet{mei2025mmragsurvey} synthesize definitions and components with an emphasis on cross-modal alignment; \citet{zheng2025ragandunderstandinginvision} bridge RAG with visual understanding and generation and discuss embodied settings;
and \citet{gao2025scalingbeyondcontext_mragsurvey} review multimodal RAG approaches for document understanding and compile a collection of multimodal RAG datasets.
Additionally, \citet{ding2025deeplearningbasedvisually} provide a comprehensive overview of deep learning–based VRD.
Compared with these works, our survey narrows the focus to visually rich documents and contributes a role-based taxonomy for how MLLMs enter the pipeline while foregrounding practical trade-offs specific to document-centric RAG.

\section{Three Roles of MLLMs in VRD RAG}
\label{sec:roles_of_mllm_in_vrd_rag}

We introduce the \emph{Emergent Large-Scale Paradigm} of multimodal RAG: the systematic use of MLLMs to move beyond text-only pipelines by treating page images, layout, and visual structure as first-class retrieval signals.
Rather than a single recipe, this paradigm appears in practice through three complementary roles that MLLMs can play in VRD pipelines:
\emph{Modality-Unifying Captioners}, which translate non-text elements into textual surrogates for conventional indexing;
\emph{Multimodal Embedders}, which map heterogeneous inputs into a shared representation space for cross-modal search;
and \emph{End-to-End Representers}, which encode whole pages directly without explicit OCR or layout parsing.
Viewing the literature through these roles provides a concrete basis for analyzing retrieval granularity, information fidelity, and system cost in \S\ref{sec:trade_offs}.

\subsection{MLLMs as Modality-Unifying Captioners}
\label{subsec:mllm-as-captioner}

As sketched in Figure~\ref{fig:main-fig} (left), this role converts non-textual elements into textual surrogates for conventional indexing and retrieval.
In the \emph{Modality-Unifying Captioner} role, systems translate non-textual inputs into textual surrogates so that retrieval and generation can proceed in the \emph{text} modality.
For VRDs, this typically means (i) OCR- and layout-aware textualization of pages and regions, and (ii) higher-level natural-language descriptions that summarize figures, tables, and UI screenshots.
The resulting text is embedded with standard text encoders and indexed alongside native document text, enabling drop-in multimodal support for existing text-only RAG stacks.

\paragraph{From captioning to document textualization}
Early captioners established language as a universal interface for vision.
\citet{vinyals2015showandtell} and \citet{xu2015showattendandtell} demonstrated global and attention-grounded image descriptions; \citet{johnson2016densecap} introduced region-level captions, inspiring fine-grained retrieval in VRDs, where figure panels or table regions should be independently retrievable. OCR-aware captioning such as TextCaps \cite{sidorov2020textcaps} explicitly reads in-image text, crucial for charts and slides where on-image text encodes semantics.
In VRD pipelines, LayoutLM \cite{xu2020layoutlm, xu2021layoutlmv2, huang2022layoutlmv3} unified OCR tokens with 2D coordinates for forms and invoices, while the DocVQA \cite{mathew2021docvqa} benchmark standardized OCR-first evaluation.
Beyond OCR, Donut \cite{kim2022donut} mapped document images directly to target text to reduce error propagation, and Pix2Struct \cite{lee2022pix2struct} turned UI/web screenshots into simplified HTML, making both approaches practical captioners that emit structured proxies well-suited for text indexing.

\paragraph{Captions as textual proxies}
The same \emph{text proxy} pattern recurs across modalities and offers lessons for VRDs.
In video and audio, \citet{miech2019howto100m} leveraged narration transcripts for supervision; \citet{xu2021videoclip} aligned video with text via contrastive pretraining; \citet{lei2018tvqa} operationalized subtitle-centric QA with temporal localization.
Error cascades in speech-to-text QA were documented by Spoken SQuAD \cite{li2018spokensquad}, underscoring the brittleness of ASR-first pipelines.
For environmental audio, AudioCaps \cite{kim2019audiocaps} and Clotho \cite{drossos2019clotho} showed that textual captions are effective surrogates for downstream retrieval and clustering.
For structured vision, \citet{johnson2015imageretrievalusingscenegraphs} treated a scene graph as a structured caption to drive semantic retrieval.
In clinical imaging, R2Gen \cite{chen2020r2gen} cast images into long textual reports, indexable as evidence in text-first RAG.

\paragraph{Practical Deployments}
Production systems generally follow a consistent approach to captioning.
They first introduce an upstream captioning layer that generates page or region summaries, verbalizes tables, and produces figure descriptions.
After this conversion stage, a mature text retriever and reader are applied to the resulting text. Practical tutorials by \citet{langchain2023multimodal} and \citet{nvidia2024multimodal} describe this convert-first-then-index workflow for slide and PDF question answering. Comparable industrial deployments and discussions of potential limitations are provided by \citet{riedler2024beyondtext}.
Evidence from these studies suggests that using stronger captioners consistently improves recall and answer quality, even when the retrieval model remains unchanged.

\paragraph{Video-RAG as a mirror for VRDs}
The captioning approach can be extended to long videos by treating time-aligned text as the primary index.
Recent systems refine this concept by explicitly captioning long videos through the extraction of automatic speech recognition (ASR), optical character recognition (OCR), and object detection outputs, which are converted into textual fragments used as retrievable evidence.
Video-RAG \cite{luo2024videorag} represents one such example, where ASR, OCR, and detection results are transformed into retrievable text aligned with sampled video frames.
SceneRAG \cite{zeng2025scenerag} incorporates ASR transcripts with timestamps and scene segmentation, together with a scene-level knowledge graph to enable multi-hop retrieval.
VideoAgent \cite{fan2024videoagent} further unifies these modalities into a memory structure that combines two-second subtitle segments with tables describing object states.
Subsequent research, including works by \citet{ren2025videorag} and \citet{jeong2025videorag}, extends these ideas to very long videos by advocating dual-channel architectures that preserve both textual proxies and visual context. This approach mirrors the best practices established in VRD tasks, where captions are paired with region crops to improve reranking and grounding.
Resources devoted to video chaptering, such as Chapter-Llama \cite{Ventura2025chapterllama} and the VidChapters-7M dataset \cite{yang2023vidchapters7m}, illustrate how ASR transcripts combined with visual features can yield robust segment-level indices.
These insights are directly applicable to VRD pipelines, where similar methods can strengthen section- or figure-level retrieval.

\paragraph{Conversion to a dominant modality}
Outside text, \emph{proxy conversion} is a common way to reuse strong tooling.
In 3D perception, MV3D \cite{chen2017mv3d}, PIXOR \cite{Yang2018pixor}, and PointPillars \cite{Lang2019PointPillars} project LiDAR point clouds to BEV or pseudo-images to leverage 2D detectors and infra.
While their target modality is vision, the strategy is analogous to VRD captioners: convert heterogeneous inputs into the most mature stack.
For enterprise VRDs, the most mature stack is text retrieval, hence captioning and structural textualization is the natural endpoint.

\paragraph{Where modern MLLMs fit}
Modern MLLMs enable seamless integration of captioning within VRD pipelines. These models generate both page-level and region-level descriptions, convert chart and table content into text by describing units, axes, trends, and outliers, and can even produce structured representations in formats such as HTML, Markdown, or JSON, following the design principles of systems like Pix2Struct and Donut.
Empirical evidence shows that employing more capable captioners leads to measurable gains in recall and answer accuracy for slide and PDF question answering tasks \cite{langchain2023multimodal, riedler2024beyondtext}, even when the retrieval process at query time remains entirely text-based.

\paragraph{Advantages}
Across modalities, the strategy of first converting heterogeneous inputs into a dominant or well-supported modality reflects a shared set of motivations.
By transforming diverse signals into text, practitioners can leverage decades of progress in indexing, retrieval, and evaluation.
Cascaded architectures allow modular replacement and incremental upgrading of components such as OCR or retrievers. This design enhances interpretability, facilitates debugging, and eases deployment and optimization in production environments. Additionally, online query latency remains unaffected, as all processing of charts and tables by VLMs is confined to the preprocessing stage.

\paragraph{Disadvantages}
Despite these advantages, the paradigm carries inherent risks.
Captioning or transcription inevitably compresses the source signal, risking the omission of fine-grained visual or temporal information. Highly structured visual elements, such as charts, diagrams, or tables, often lose numerical precision or relational cues when summarized in free-form text, a weakness that motivates corrective research like chart-to-text generation.
Recognition errors from ASR or OCR can cascade, significantly degrading downstream retrieval or QA accuracy. The Spoken-SQuAD dataset quantified this impact for speech QA \cite{li2018spokensquad}, while models such as Donut explicitly sought to eliminate OCR-induced error chains through end-to-end document decoding \cite{kim2022donut}.
Furthermore, preprocessing for large-scale captioning with this approach can be costly.
Processing vast repositories of documents, each potentially containing numerous images and tables, requires substantial computational resources and time for the MLLM to generate descriptions for every non-textual element. This upfront cost can be a major bottleneck, especially for dynamic datasets where new multimodal content is frequently added.

This suggests that while the modality-unifying captioner role offers an accessible path to multimodal RAG, it may be best suited for applications where the non-textual elements are relatively simple, where some information loss is tolerable, or where the scale of data does not make preprocessing costs insurmountable.

\subsection{MLLMs as Multimodal Embedders}
\label{subsec:mllm-as-mmemb}

As shown in Figure~\ref{fig:main-fig} (middle), this role embeds heterogeneous inputs into a shared space to enable cross-modal search and matching.

While the captioner role (\S\ref{subsec:mllm-as-captioner}) is practical, restricting MLLMs solely to text conversion has inherent limitations.
In response to these constraints, the research community has increasingly focused on leveraging the advanced representation capabilities of MLLMs to enhance multimodal RAG.
A prominent direction within this effort involves utilizing MLLMs as \emph{Multimodal Embedders}.
In this role, MLLMs function directly as powerful embedding models, transforming data from diverse modalities into a shared, rich semantic feature space.

\paragraph{The Core Mechanism}
Instead of converting modalities to text, the MLLM learns to map inputs from different modalities into a common high-dimensional vector space.
In this shared space, the embeddings of semantically related items from different modalities are expected to be close to each other, allowing for direct comparison, similarity search, and retrieval across modalities.
For example, an image query could retrieve relevant textual passages, or a textual query could retrieve relevant images and text.

\paragraph{Historical Roots}
This fundamental idea, unifying disparate modalities into a common representational space to facilitate joint reasoning and retrieval, has deep historical roots.
The most canonical instantiation is perhaps CLIP \cite{radford2021clip} and its numerous successors \cite{jia2021ALIGN, Zhai2022LiT, li2022blip, li2023blip2, zhai2023siglip, yao2021filip, yu2022coca}, which align text and image representations via contrastive learning, establishing CLIP as the de facto standard embedding backbone in early multimodal RAG systems.
Earlier precursors include DeViSE \cite{frome2013devise}, which projected visual features into word2vec semantic space for zero-shot recognition; Deep CCA \cite{andrew2013deepcca} and its deep extensions DGCCA \cite{benton2019dgcca}, which learned shared subspaces via canonical correlation analysis; VSE++ \cite{faghri2018vsepp}, which emphasized hard negative mining for improved alignment; and SCAN \cite{Lee2018SCAN}, which introduced stacked cross-attention to enable fine-grained word-region alignment for stronger image–text matching.
More recently, ImageBind \cite{Girdhar2023ImageBind} unified six modalities into a single embedding space, achieving cross-modal alignment using only image-paired data as a bridging signal.

\paragraph{The Shift to MLLM-based Embedders}
Nevertheless, recent studies \cite{zhou2024vista} have indicated that the text embedding capabilities of these vision-language models (e.g., CLIP) are comparatively inferior to those of specialized text embedding models.
This limitation may hinder their effectiveness in tasks involving text-intensive multimodal documents. Drawing inspiration from pioneering work such as LLM2Vec \cite{behnamghader2024llm2vec}, there has been a surge in research efforts aimed at repurposing MLLMs as embedding models \cite{jiang2025e5v, jiang2025vlm2vec, meng2025vlm2vecv2, zhou2024mmret, lin2025mmembed, zhang2025gme, lan2025llave, liu2024lamra, chen2025moca, lee2025GeneralizedContrastivLelearning, lin2025VC2L}.
Representative approaches include:
VLM2Vec \cite{jiang2025vlm2vec, meng2025vlm2vecv2}, which endows VLMs with instruction-aware embedding capabilities and reports consistent gains across the Massive Multimodal Embedding Benchmark (MMEB);
MM-Embed \cite{lin2025mmembed}, which identifies and mitigates modality bias via modality-aware hard negative mining and continual text-to-text fine-tuning; and
E5-V \cite{jiang2025e5v}, which surprisingly achieves state-of-the-art multimodal retrieval by training exclusively on text pairs, leveraging prompting to bridge modalities and drastically reduce annotation and training costs.

\paragraph{Training Strategy}
This methodology, mirroring LLM2Vec, transforms MLLMs into CLIP-analogous representation models by embedding diverse modalities into a shared feature space.
For instance, in the case of VLMs, the process involves aggregating extensive datasets similar to CLIP's training corpus.
During training, the EOS token serves as the representative token, and contrastive learning employs InfoNCE loss \cite{oord2019infonce}.
Through this approach, textual and visual modalities are seamlessly integrated into a unified feature space.
Leveraging the MLLM's world knowledge from multimodal next-token prediction \cite{chen2024tokenpredictionmultimodalintelligence}, this method demonstrates exceptional representational capacity across diverse data types.
Moreover, it offers the flexibility to replace existing embedding models with minimal disruption.

In addition to CLIP-style training, MoCa \cite{chen2025moca} converts causal VLMs into bidirectional multimodal embedders via continual pretraining and heterogeneous contrastive finetuning.
Vision-centric contrastive learning (VC$^2$L) \cite{lin2025VC2L} renders mixed text–image content into pixels to avoid OCR misalignment.
General training advances include a generalized contrastive loss (GCL) \cite{lee2025GeneralizedContrastivLelearning} that jointly contrasts text, image, and fused representations within a batch.

\paragraph{The Role of Data and Synthetic Supervision}
Beyond algorithmic design, data composition and synthetic supervision play pivotal roles.
\citet{zhang2025gme} target universal multimodal retrieval over visually rich documents, emphasizing balanced modality mixing and efficient generation of fused-modality training pairs.
\citet{zhou2024mmret} scale synthetic supervision by generating instruction-style queries over image pairs, significantly enhancing zero-shot generalization.

\paragraph{Optimization}
On the optimization front, LLaVE \cite{lan2025llave} introduces hardness-weighted contrastive learning to better separate ambiguous negatives.
Complementing standard bi-encoder frameworks, LamRA \cite{liu2024lamra} attaches lightweight LoRA \cite{hu2022lora} heads to generative MLLMs, unifying retrieval and reranking and enabling strong transfer to unseen retrieval tasks.

\paragraph{Empirical Evidence and Performance}
The superior impact of these MLLM-derived embeddings is evident in downstream performance, with extensive evaluations confirming better representational quality and metrics.
Table~\ref{tab:mmeb_exp}, compiling results from the MMEB introduced by \citet{jiang2025vlm2vec}, illustrates this trend.
Notably, employing MLLMs as Multimodal Embedders yields substantial performance enhancements compared to RAG systems using conventional multimodal embedding models like CLIP.
In VQA-related tasks, for example, MLLMs as Multimodal Embedders leverage their inherent advanced visual reasoning capabilities, highlighting their distinct advantage.

\paragraph{Complementary advances in reranking}
Several concurrent efforts investigate VLM-based reranking methods to complement retrieval with reasoning-aware relevance modeling \cite{xu2025mmr5, wasserman2025DocReRank, chen2025cmrag, gong2025MHierRAG}.

\subsection{MLLMs as End-to-End Representers}
\label{subsec:mllm-as-e2erepr}

As illustrated in Figure~\ref{fig:main-fig} (right), the \emph{End-to-End Representers} role encodes whole pages directly to retrieve at page granularity.
It uses MLLMs to generate holistic representations directly from entire multimodal inputs, such as treating full document pages as single images.
Instead of breaking a document down into its constituent parts and processing them separately or converting them, the MLLM takes a more holistic view.
For example, an entire PDF page, with its complex layout of text, images, and graphical elements, might be fed as a single image input to the MLLM, which then generates a unified representation for that entire page.

A key characteristic of this approach is that it bypasses intermediate steps like explicit OCR or layout parsing.
Traditional document processing pipelines often rely on separate modules for OCR, layout analysis, and then subsequent processing of these extracted elements. Each of these stages can introduce errors.

\paragraph{Rationale}
To illustrate this methodology, consider the example of VLMs. In this instance, specific components within traditional RAG pipelines are replaced with VLMs, thereby enabling direct end-to-end representation generation.
This approach is motivated by two key factors.
Firstly, previous research has demonstrated that the process of OCR introduces noise into RAG systems, degrading their performance \cite{zhang2025ocrhindersrag, xie2025TextlessRAG}.
Secondly, the advanced visual comprehension capabilities of contemporary VLMs render separate identification of layouts, tables, images, and other discrete elements unnecessary. Instead, an entire PDF page can be treated as a single image input to a VLM, thereby facilitating the production of a holistic representation.

\paragraph{Exemplary Models}
Significant contributions have been made in this domain, with DSE \cite{ma2024dse}, ColPali \cite{faysse2025colpali} and VisRAG \cite{yu2025visrag} being particularly noteworthy examples.
DSE has the capacity to convert document screenshots directly into dense vectors for retrieval.
ColPali incorporates the late-interaction matching mechanism of ColBERT \cite{Khattab2020colbert}, embedding document page images into high-dimensional vector spaces for retrieval.
This method excels at capturing intricate visual details and is simple, fast, and end-to-end trainable.
Similarly, VisRAG directly encodes and retrieves document pages, mitigating information loss while fully exploiting the visual content present in documents.
These approaches adopt InfoNCE loss for training, aligning with the training approach of the \emph{Multimodal Embedders} role.

\paragraph{Beyond Single-Page}
Beyond page-level late-interaction encoders, multi-page representers decouple retrieval and reasoning.
DREAM \cite{zhang2025dream} integrates hierarchical multimodal retrieval and a multi-page VLM with global and token-level cross-page attention.
ColMate \cite{masry2025colmate}, while primarily an embedder, inherits ColBERT-style end-to-end matching over page images and masked text.
Industry efforts such as Docopilot \cite{Duan2025Docopilot} study non-RAG, end-to-end multi-round document understanding over the Doc-750K corpus, complementary to retrieval-centric approaches.

\paragraph{Advantage}
This end-to-end Representer methodology capitalizes on the advanced representational capabilities of MLLMs while concurrently reducing the overall latency of the pipeline \cite{faysse2025colpali, yu2025visrag}.
In traditional multimodal RAG, predominant latency sources are initial layout analysis, segmentation, and OCR, not embedding itself.
Employing MLLMs for end-to-end recognition, despite a slight increase in embedding duration, results in a substantial reduction in total processing time.
This is demonstrated in Table~\ref{tab:latency_comparison}, which compares the latency of an OCR-reliant pipeline with an MLLM-based end-to-end representer, showing a reduction in total offline latency for the MLLM-based approach due to the elimination of parsing overhead.

This approach can also reduce the noise caused by imperfect parsing.
OCR errors, misinterpretations of document layout, or failures to correctly segment different content blocks can degrade the quality of information fed into a RAG system.
An MLLM that directly sees the entire page might learn to be more robust to such variations or low-quality inputs, as it can leverage the global context of the page.
This holistic processing can be particularly advantageous for ingesting large volumes of complex documents, such as scanned PDFs or documents with unconventional layouts, where traditional parsing tools might struggle.

Furthermore, end-to-end training integrates the previously acquired world knowledge and inherent capabilities of MLLMs, and thus elevates the performance ceiling of multimodal RAG systems.

\section{Trade-offs and Future Directions}
\label{sec:trade_offs}

While integrating MLLMs into RAG systems offers significant benefits, this paradigm also presents challenges and is not universally optimal. Key limitations involve retrieval granularity, information fidelity, and computational and storage demands.

\subsection{Retrieval Granularity and Interpretability}
\label{subsec:retrieval_granularity_and_interpretability}

\subsubsection{Coarse vs. Fine-Grained Retrieval}

The \emph{End-to-End Representer} role, despite preprocessing benefits, often yields coarser retrieval granularity.
For example, both ColPali and VisRAG adopt the page as the retrieval unit \cite{faysse2025colpali, yu2025visrag}.
While representing a whole page with one vector identifies relevant pages, it obscures fine-grained details, forcing a secondary search for specific facts, unlike text-based RAG systems that retrieve individual paragraphs or sentences.
This highlights a fundamental tension: holistic processing improves robustness but sacrifices retrieval precision, whereas fine-grained retrieval enhances precision but risks losing global context or suffering from error propagation.

\subsubsection{Information Loss in Conversion}
Similarly, the Modality-Unifying Captioner role, which converts non-textual elements to text, inherently suffers from information loss, as textual descriptions rarely capture the full richness of images, tables, or diagrams.
This preprocessing information loss directly degrades fidelity: if the LLM generator receives incomplete or oversimplified context, the final output will lack nuance and accuracy, undermining the RAG system's purpose.

Ideal granularity and acceptable information loss are application-dependent.
For instance, general summarization may tolerate coarser granularity, whereas fact-checking demands high-fidelity, fine-grained retrieval. This tension highlights the need for adaptive, task-aware systems rather than a single, universally optimal strategy.

Recent studies \cite{gong2025MHierRAG, chen2025cmrag, zhang2025dream} further demonstrate that adaptive hierarchical and co-modality retrieval strategies can effectively recover fine-grained evidence and improve cross-page reasoning in visually rich documents.

\subsection{Computational Overhead and Costs}
\label{subsec:computational_overhead_and_costs}

\subsubsection{Increased Latency}

Generating rich multimodal embeddings or detailed textual captions using MLLMs is computationally intensive.
Figure~\ref{fig:retrieval-performance} shows MLLMs as multimodal embedders incur substantially higher latency during both offline encoding and online searching compared to CLIP-based models.
This starkly illustrates why model miniaturization is essential for broader applicability.

\subsubsection{Substantial Storage Demands}

MLLM-RAG systems also face substantial storage demands.
MLLMs in the \emph{Multimodal Embedder} role produce high-dimensional embeddings, often significantly larger than traditional text vectors, to capture rich cross-modal information.
Storing these vectors for large corpora can become prohibitive.
For instance, \citet{lin2025mmembed} report that its index storage demands exceed those of CLIP-based models by a factor of five or more.

This increased storage footprint not only incurs direct hardware costs but also degrades efficiency by slowing index loading and vector searches, compounding latency issues.

Potential solutions include model miniaturization via higher-quality data or knowledge distillation \cite{hinton2015distillingknowledgeneuralnetwork}, which could produce compact Multimodal Small Language Models to address these root challenges.

Another promising avenue is the adoption of a Matryoshka-style multimodal learning framework \cite{sturua2024jinaembeddingsv3, cai2025matryoshka}, which learns representations across multiple granularities. By dynamically selecting inference modes, this approach could offer a scalable performance-cost gradient tailored to downstream tasks.

Recent works \cite{rajendran2025EcoDoc, yan2025docpruner, gunther2025jinaembeddingsv4, masry2025colmate} have also explored efficiency-oriented solutions that balance accuracy and cost through adaptive routing, vector pruning, and lightweight embedding designs.

\subsection{Challenges in Evaluation Metrics}
\label{subsec:evaluation_metrics_challenges}

Evaluating multimodal RAG remains fundamentally difficult because traditional metrics, largely developed for text-only settings, cannot fully capture the fidelity and interpretability of cross-modal reasoning.
While frameworks such as RAGAs \cite{es-etal-2024-ragas} and ARES \cite{saadfalcon2024ares} provide initial measures for faithfulness and relevance, multimodal scenarios introduce new failure sources, including misaligned visual grounding and inconsistencies between retrieved and generated evidence \cite{mortaheb2025ragcheckevaluatingmultimodalretrieval}.
Recent benchmarks \cite{wasserman2025realmmrag, peng2025UniDocBench} highlight that current systems often underperform on real-world, document-heavy, and paraphrase-variant data, underscoring a persistent gap between laboratory metrics and practical robustness.
Human-centered datasets can also help narrow this pragmatic gap \cite{zhang2025auradial}.

A more holistic evaluation paradigm is needed, combining end-to-end performance with modality-aware diagnostics such as table and figure grounding accuracy, cross-page evidence localization, and paraphrase robustness, aligning with broader calls for benchmarks that prioritize safety and real-world user needs \cite{zhang2025escalatorproblem}.
Progress in this direction will enable fairer comparison across retrieval granularity levels and provide actionable signals for improving factual alignment and interpretability in visually rich document RAG systems.

\section{Conclusion}
\label{sec:conclusion}

This survey has chartered the evolving landscape of Retrieval-Augmented Generation for visually rich documents, focusing on the critical roles played by MLLMs. We have structured this emergent field by proposing a taxonomy of three primary roles: \emph{Modality-Unifying Captioners}, \emph{Multimodal Embedders}, and \emph{End-to-End Representers}.

Our analysis reveals that there is no single, universally optimal solution. Instead, practitioners face a distinct set of trade-offs.
The \emph{Captioner} role offers a pragmatic path to multimodal support by integrating with mature, text-based RAG pipelines, but at the risk of information loss and error cascades from imperfect textual conversion.
The \emph{Embedder} role enables true cross-modal search by unifying modalities in a shared vector space, but this power often comes at the cost of significant computational and storage overhead.
Finally, the \emph{Representer} role provides robustness by bypassing brittle OCR and parsing steps, but this simplicity typically sacrifices retrieval precision by operating at a coarse, page-level granularity.

These findings highlight a tension in the field: a balancing act between retrieval granularity, information fidelity, computational cost, and pipeline simplicity.
As the field matures, we anticipate future research will focus on three challenges.
First, the development of adaptive and hierarchical retrieval methods to dynamically blend coarse-grained and fine-grained retrieval to get the best of both.
Second, the need for model miniaturization and efficiency, producing smaller, faster MLLMs that make these advanced techniques practical for real-world latency and storage budgets.
Finally, the design of next-generation evaluation benchmarks that move beyond simple text-based metrics to holistically measure factual accuracy, cross-modal grounding, and the interpretability of RAG systems handling complex, visually-grounded evidence.

\section*{Limitations}
\label{sec:limitations}

This survey has limitations.
Firstly, its scope is constrained by available literature on MLLMs in multimodal RAG. The generalizability of the synthesized findings may be limited by the datasets, MLLMs, and tasks predominantly featured in these studies.
Secondly, while performance and latency are discussed based on reported figures, this survey does not account for the variability in hardware configurations or deployment environments used in those studies, which could impact real-world applicability comparisons.
Lastly, the reviewed literature often focuses more on technical and performance aspects, with less emphasis on user-centric evaluation metrics such as nuanced interpretability and usability.
This survey reflects that focus, leaving broader user-centric analyses for future work or dedicated studies.

\bibliography{main}

@inproceedings{lewis2020rag,
  author       = {Patrick S. H. Lewis and
                  Ethan Perez and
                  Aleksandra Piktus and
                  Fabio Petroni and
                  Vladimir Karpukhin and
                  Naman Goyal and
                  Heinrich K{\"{u}}ttler and
                  Mike Lewis and
                  Wen{-}tau Yih and
                  Tim Rockt{\"{a}}schel and
                  Sebastian Riedel and
                  Douwe Kiela},
  title        = {Retrieval-Augmented Generation for Knowledge-Intensive {NLP} Tasks},
  booktitle    = {Advances in Neural Information Processing Systems 33: Annual Conference
                  on Neural Information Processing Systems 2020, NeurIPS 2020, December
                  6-12, 2020, virtual},
  year         = {2020},
  url          = {https://proceedings.neurips.cc/paper/2020/hash/6b493230205f780e1bc26945df7481e5-Abstract.html}
}

@misc{gao2024ragforllms_survey,
      title={Retrieval-Augmented Generation for Large Language Models: A Survey}, 
      author={Yunfan Gao and Yun Xiong and Xinyu Gao and Kangxiang Jia and Jinliu Pan and Yuxi Bi and Yi Dai and Jiawei Sun and Meng Wang and Haofen Wang},
      year={2024},
      eprint={2312.10997},
      archivePrefix={arXiv},
      primaryClass={cs.CL},
      url={https://arxiv.org/abs/2312.10997}, 
}

@inproceedings{fan2024asurveyonragmeetingllms,
    author = {Fan, Wenqi and Ding, Yujuan and Ning, Liangbo and Wang, Shijie and Li, Hengyun and Yin, Dawei and Chua, Tat-Seng and Li, Qing},
    title = {A Survey on RAG Meeting LLMs: Towards Retrieval-Augmented Large Language Models},
    year = {2024},
    isbn = {9798400704901},
    publisher = {Association for Computing Machinery},
    address = {New York, NY, USA},
    url = {https://doi.org/10.1145/3637528.3671470},
    doi = {10.1145/3637528.3671470},
    booktitle = {Proceedings of the 30th ACM SIGKDD Conference on Knowledge Discovery and Data Mining},
    pages = {6491–6501},
    numpages = {11},
    location = {Barcelona, Spain},
    series = {KDD '24}
}

@inproceedings{shuster2021ragreduceshallucination,
  author       = {Kurt Shuster and
                  Spencer Poff and
                  Moya Chen and
                  Douwe Kiela and
                  Jason Weston},
  title        = {Retrieval Augmentation Reduces Hallucination in Conversation},
  booktitle    = {Findings of the Association for Computational Linguistics: {EMNLP}
                  2021, Virtual Event / Punta Cana, Dominican Republic, 16-20 November,
                  2021},
  year         = {2021},
  url          = {https://doi.org/10.18653/v1/2021.findings-emnlp.320}
}

@misc{mao2021generationaugmentedretrievalopendomainquestion,
      title={Generation-Augmented Retrieval for Open-domain Question Answering}, 
      author={Yuning Mao and Pengcheng He and Xiaodong Liu and Yelong Shen and Jianfeng Gao and Jiawei Han and Weizhu Chen},
      year={2021},
      eprint={2009.08553},
      archivePrefix={arXiv},
      primaryClass={cs.CL},
      url={https://arxiv.org/abs/2009.08553}, 
}

@article{siriwardhana2023improvingdomainadaptionofrag,
  author       = {Shamane Siriwardhana and
                  Rivindu Weerasekera and
                  Tharindu Kaluarachchi and
                  Elliott Wen and
                  Rajib Rana and
                  Suranga Nanayakkara},
  title        = {Improving the Domain Adaptation of Retrieval Augmented Generation
                  {(RAG)} Models for Open Domain Question Answering},
  journal      = {Trans. Assoc. Comput. Linguistics},
  volume       = {11},
  pages        = {1--17},
  year         = {2023},
  url          = {https://doi.org/10.1162/tacl\_a\_00530},
  doi          = {10.1162/TACL\_A\_00530},
  bibsource    = {dblp computer science bibliography, https://dblp.org}
}

@misc{thulke2021efficientretrievalaugmentedgeneration,
      title={Efficient Retrieval Augmented Generation from Unstructured Knowledge for Task-Oriented Dialog}, 
      author={David Thulke and Nico Daheim and Christian Dugast and Hermann Ney},
      year={2021},
      eprint={2102.04643},
      archivePrefix={arXiv},
      primaryClass={cs.CL},
      url={https://arxiv.org/abs/2102.04643}, 
}

@inproceedings{chen2022murag,
  author       = {Wenhu Chen and
                  Hexiang Hu and
                  Xi Chen and
                  Pat Verga and
                  William W. Cohen},
  editor       = {Yoav Goldberg and
                  Zornitsa Kozareva and
                  Yue Zhang},
  title        = {MuRAG: Multimodal Retrieval-Augmented Generator for Open Question
                  Answering over Images and Text},
  booktitle    = {Proceedings of the 2022 Conference on Empirical Methods in Natural
                  Language Processing, {EMNLP} 2022, Abu Dhabi, United Arab Emirates,
                  December 7-11, 2022},
  pages        = {5558--5570},
  publisher    = {Association for Computational Linguistics},
  year         = {2022},
  url          = {https://doi.org/10.18653/v1/2022.emnlp-main.375},
  doi          = {10.18653/V1/2022.EMNLP-MAIN.375},
  bibsource    = {dblp computer science bibliography, https://dblp.org}
}

@misc{yasunaga2023retrievalaugmentedmultimodallanguagemodeling,
      title={Retrieval-Augmented Multimodal Language Modeling}, 
      author={Michihiro Yasunaga and Armen Aghajanyan and Weijia Shi and Rich James and Jure Leskovec and Percy Liang and Mike Lewis and Luke Zettlemoyer and Wen-tau Yih},
      year={2023},
      eprint={2211.12561},
      archivePrefix={arXiv},
      primaryClass={cs.CV},
      url={https://arxiv.org/abs/2211.12561}, 
}

@InProceedings{ramesh2021dalle,
  title = 	 {Zero-Shot Text-to-Image Generation},
  author =       {Ramesh, Aditya and Pavlov, Mikhail and Goh, Gabriel and Gray, Scott and Voss, Chelsea and Radford, Alec and Chen, Mark and Sutskever, Ilya},
  booktitle = 	 {Proceedings of the 38th International Conference on Machine Learning},
  pages = 	 {8821--8831},
  year = 	 {2021},
  editor = 	 {Meila, Marina and Zhang, Tong},
  volume = 	 {139},
  series = 	 {Proceedings of Machine Learning Research},
  month = 	 {18--24 Jul},
  publisher =    {PMLR},
  url = 	 {https://proceedings.mlr.press/v139/ramesh21a.html},
}

@inproceedings{wei2024uniir,
  author       = {Cong Wei and
                  Yang Chen and
                  Haonan Chen and
                  Hexiang Hu and
                  Ge Zhang and
                  Jie Fu and
                  Alan Ritter and
                  Wenhu Chen},
  editor       = {Ales Leonardis and
                  Elisa Ricci and
                  Stefan Roth and
                  Olga Russakovsky and
                  Torsten Sattler and
                  G{\"{u}}l Varol},
  title        = {UniIR: Training and Benchmarking Universal Multimodal Information
                  Retrievers},
  booktitle    = {Computer Vision - {ECCV} 2024 - 18th European Conference, Milan, Italy,
                  September 29-October 4, 2024, Proceedings, Part {LXXXVII}},
  series       = {Lecture Notes in Computer Science},
  volume       = {15145},
  pages        = {387--404},
  publisher    = {Springer},
  year         = {2024},
  url          = {https://doi.org/10.1007/978-3-031-73021-4\_23}
}

@misc{kim2025geniusgenerativeframeworkuniversal,
      title={GENIUS: A Generative Framework for Universal Multimodal Search}, 
      author={Sungyeon Kim and Xinliang Zhu and Xiaofan Lin and Muhammet Bastan and Douglas Gray and Suha Kwak},
      year={2025},
      eprint={2503.19868},
      archivePrefix={arXiv},
      primaryClass={cs.IR},
      url={https://arxiv.org/abs/2503.19868}, 
}

@misc{bi2025describedwordssimpleunified,
      title={Everything Can Be Described in Words: A Simple Unified Multi-Modal Framework with Semantic and Temporal Alignment}, 
      author={Xiaowei Bi and Zheyuan Xu},
      year={2025},
      eprint={2503.09081},
      archivePrefix={arXiv},
      primaryClass={cs.CV},
      url={https://arxiv.org/abs/2503.09081}, 
}

@misc{zhao2023retrievingmultimodalinformationaugmented,
      title={Retrieving Multimodal Information for Augmented Generation: A Survey}, 
      author={Ruochen Zhao and Hailin Chen and Weishi Wang and Fangkai Jiao and Xuan Long Do and Chengwei Qin and Bosheng Ding and Xiaobao Guo and Minzhi Li and Xingxuan Li and Shafiq Joty},
      year={2023},
      eprint={2303.10868},
      archivePrefix={arXiv},
      primaryClass={cs.CL},
      url={https://arxiv.org/abs/2303.10868}, 
}

@inproceedings{liu2023llava,
    title={Visual Instruction Tuning},
    author={Haotian Liu and Chunyuan Li and Qingyang Wu and Yong Jae Lee},
    booktitle={Thirty-seventh Conference on Neural Information Processing Systems},
    year={2023},
    url={https://openreview.net/forum?id=w0H2xGHlkw}
}

@misc{liu2024llavanext,
    title={LLaVA-NeXT: Improved reasoning, OCR, and world knowledge},
    url={https://llava-vl.github.io/blog/2024-01-30-llava-next/},
    author={Liu, Haotian and Li, Chunyuan and Li, Yuheng and Li, Bo and Zhang, Yuanhan and Shen, Sheng and Lee, Yong Jae},
    month={January},
    year={2024}
}

@misc{bai2023qwenvl,
      title={Qwen-VL: A Versatile Vision-Language Model for Understanding, Localization, Text Reading, and Beyond}, 
      author={Jinze Bai and Shuai Bai and Shusheng Yang and Shijie Wang and Sinan Tan and Peng Wang and Junyang Lin and Chang Zhou and Jingren Zhou},
      year={2023},
      eprint={2308.12966},
      archivePrefix={arXiv},
      primaryClass={cs.CV},
      url={https://arxiv.org/abs/2308.12966}, 
}

@misc{wang2024qwen2vl,
      title={Qwen2-VL: Enhancing Vision-Language Model's Perception of the World at Any Resolution}, 
      author={Peng Wang and Shuai Bai and Sinan Tan and Shijie Wang and Zhihao Fan and Jinze Bai and Keqin Chen and Xuejing Liu and Jialin Wang and Wenbin Ge and Yang Fan and Kai Dang and Mengfei Du and Xuancheng Ren and Rui Men and Dayiheng Liu and Chang Zhou and Jingren Zhou and Junyang Lin},
      year={2024},
      eprint={2409.12191},
      archivePrefix={arXiv},
      primaryClass={cs.CV},
      url={https://arxiv.org/abs/2409.12191}, 
}

@misc{bai2025qwen2_5vl,
      title={Qwen2.5-VL Technical Report}, 
      author={Shuai Bai and Keqin Chen and Xuejing Liu and Jialin Wang and Wenbin Ge and Sibo Song and Kai Dang and Peng Wang and Shijie Wang and Jun Tang and Humen Zhong and Yuanzhi Zhu and Mingkun Yang and Zhaohai Li and Jianqiang Wan and Pengfei Wang and Wei Ding and Zheren Fu and Yiheng Xu and Jiabo Ye and Xi Zhang and Tianbao Xie and Zesen Cheng and Hang Zhang and Zhibo Yang and Haiyang Xu and Junyang Lin},
      year={2025},
      eprint={2502.13923},
      archivePrefix={arXiv},
      primaryClass={cs.CV},
      url={https://arxiv.org/abs/2502.13923}, 
}

@InProceedings{chen2024internvl,
    author    = {Chen, Zhe and Wu, Jiannan and Wang, Wenhai and Su, Weijie and Chen, Guo and Xing, Sen and Zhong, Muyan and Zhang, Qinglong and Zhu, Xizhou and Lu, Lewei and Li, Bin and Luo, Ping and Lu, Tong and Qiao, Yu and Dai, Jifeng},
    title     = {InternVL: Scaling up Vision Foundation Models and Aligning for Generic Visual-Linguistic Tasks},
    booktitle = {Proceedings of the IEEE/CVF Conference on Computer Vision and Pattern Recognition (CVPR)},
    month     = {June},
    year      = {2024},
    pages     = {24185-24198}
}

@misc{chen2024internvl1_5,
      title={How Far Are We to GPT-4V? Closing the Gap to Commercial Multimodal Models with Open-Source Suites}, 
      author={Zhe Chen and Weiyun Wang and Hao Tian and Shenglong Ye and Zhangwei Gao and Erfei Cui and Wenwen Tong and Kongzhi Hu and Jiapeng Luo and Zheng Ma and Ji Ma and Jiaqi Wang and Xiaoyi Dong and Hang Yan and Hewei Guo and Conghui He and Botian Shi and Zhenjiang Jin and Chao Xu and Bin Wang and Xingjian Wei and Wei Li and Wenjian Zhang and Bo Zhang and Pinlong Cai and Licheng Wen and Xiangchao Yan and Min Dou and Lewei Lu and Xizhou Zhu and Tong Lu and Dahua Lin and Yu Qiao and Jifeng Dai and Wenhai Wang},
      year={2024},
      eprint={2404.16821},
      archivePrefix={arXiv},
      primaryClass={cs.CV},
      url={https://arxiv.org/abs/2404.16821}, 
}

@misc{chen2025internvl2_5,
      title={Expanding Performance Boundaries of Open-Source Multimodal Models with Model, Data, and Test-Time Scaling}, 
      author={Zhe Chen and Weiyun Wang and Yue Cao and Yangzhou Liu and Zhangwei Gao and Erfei Cui and Jinguo Zhu and Shenglong Ye and Hao Tian and Zhaoyang Liu and Lixin Gu and Xuehui Wang and Qingyun Li and Yimin Ren and Zixuan Chen and Jiapeng Luo and Jiahao Wang and Tan Jiang and Bo Wang and Conghui He and Botian Shi and Xingcheng Zhang and Han Lv and Yi Wang and Wenqi Shao and Pei Chu and Zhongying Tu and Tong He and Zhiyong Wu and Huipeng Deng and Jiaye Ge and Kai Chen and Kaipeng Zhang and Limin Wang and Min Dou and Lewei Lu and Xizhou Zhu and Tong Lu and Dahua Lin and Yu Qiao and Jifeng Dai and Wenhai Wang},
      year={2025},
      eprint={2412.05271},
      archivePrefix={arXiv},
      primaryClass={cs.CV},
      url={https://arxiv.org/abs/2412.05271}, 
}

@misc{zhu2025internvl3,
    title={Internvl3: Exploring advanced training and test-time recipes for open-source multimodal models},
    author={Zhu, Jinguo and Wang, Weiyun and Chen, Zhe and Liu, Zhaoyang and Ye, Shenglong and Gu, Lixin and Tian, Hao and Duan, Yuchen and Su, Weijie and Shao, Jie and others},
    year={2025},
    eprint={2504.10479},
    archivePrefix={arXiv},
    primaryClass={cs.CV},
    url={https://arxiv.org/abs/2504.10479}, 
}

@misc{wang2025internvl3_5,
    title={InternVL3.5: Advancing Open-Source Multimodal Models in Versatility, Reasoning, and Efficiency},
    author={Wang, Weiyun and Gao, Zhangwei and Gu, Lixin and Pu, Hengjun and Cui, Long and Wei, Xingguang and Liu, Zhaoyang and Jing, Linglin and Ye, Shenglong and Shao, Jie and others},
    year={2025},
    eprint={2508.18265},
    archivePrefix={arXiv},
    primaryClass={cs.CV},
    url={https://arxiv.org/abs/2508.18265},
}

@inproceedings{
    jiang2025vlm2vec,
    title={{VLM}2Vec: Training Vision-Language Models for Massive Multimodal Embedding Tasks},
    author={Ziyan Jiang and Rui Meng and Xinyi Yang and Semih Yavuz and Yingbo Zhou and Wenhu Chen},
    booktitle={The Thirteenth International Conference on Learning Representations},
    year={2025},
    url={https://openreview.net/forum?id=TE0KOzWYAF}
}

@misc{meng2025vlm2vecv2,
      title={VLM2Vec-V2: Advancing Multimodal Embedding for Videos, Images, and Visual Documents}, 
      author={Rui Meng and Ziyan Jiang and Ye Liu and Mingyi Su and Xinyi Yang and Yuepeng Fu and Can Qin and Zeyuan Chen and Ran Xu and Caiming Xiong and Yingbo Zhou and Wenhu Chen and Semih Yavuz},
      year={2025},
      eprint={2507.04590},
      archivePrefix={arXiv},
      primaryClass={cs.CV},
      url={https://arxiv.org/abs/2507.04590}, 
}

@inproceedings{
    lin2025mmembed,
    title={{MM}-{EMBED}: {UNIVERSAL} {MULTIMODAL} {RETRIEVAL} {WITH} {MULTIMODAL} {LLMS}},
    author={Sheng-Chieh Lin and Chankyu Lee and Mohammad Shoeybi and Jimmy Lin and Bryan Catanzaro and Wei Ping},
    booktitle={The Thirteenth International Conference on Learning Representations},
    year={2025},
    url={https://openreview.net/forum?id=i45NQb2iKO}
}

@inproceedings{
    lee2025nvembed,
    title={{NV}-Embed: Improved Techniques for Training {LLM}s as Generalist Embedding Models},
    author={Chankyu Lee and Rajarshi Roy and Mengyao Xu and Jonathan Raiman and Mohammad Shoeybi and Bryan Catanzaro and Wei Ping},
    booktitle={The Thirteenth International Conference on Learning Representations},
    year={2025},
    url={https://openreview.net/forum?id=lgsyLSsDRe}
}

@inproceedings{yu2025visrag,
    title={Vis{RAG}: Vision-based Retrieval-augmented Generation on Multi-modality Documents},
    author={Shi Yu and Chaoyue Tang and Bokai Xu and Junbo Cui and Junhao Ran and Yukun Yan and Zhenghao Liu and Shuo Wang and Xu Han and Zhiyuan Liu and Maosong Sun},
    booktitle={The Thirteenth International Conference on Learning Representations},
    year={2025},
    url={https://openreview.net/forum?id=zG459X3Xge}
}

@inproceedings{
    faysse2025colpali,
    title={ColPali: Efficient Document Retrieval with Vision Language Models},
    author={Manuel Faysse and Hugues Sibille and Tony Wu and Bilel Omrani and Gautier Viaud and CELINE HUDELOT and Pierre Colombo},
    booktitle={The Thirteenth International Conference on Learning Representations},
    year={2025},
    url={https://openreview.net/forum?id=ogjBpZ8uSi}
}

@inproceedings{
    behnamghader2024llm2vec,
    title={{LLM}2Vec: Large Language Models Are Secretly Powerful Text Encoders},
    author={Parishad BehnamGhader and Vaibhav Adlakha and Marius Mosbach and Dzmitry Bahdanau and Nicolas Chapados and Siva Reddy},
    booktitle={First Conference on Language Modeling},
    year={2024},
    url={https://openreview.net/forum?id=IW1PR7vEBf}
}

@misc{
    jiang2025e5v,
    title={E5-V: Universal Embeddings with Multimodal Large Language Models},
    author={Ting Jiang and Shaohan Huang and Minghui Song and Zihan Zhang and Haizhen Huang and Liang Wang and Furu Wei and Weiwei Deng and Feng Sun and Qi Zhang and deqing wang and Fuzhen Zhuang},
    year={2025},
    url={https://openreview.net/forum?id=rD6LQagatR}
}

@misc{zhou2024mmret,
      title={MegaPairs: Massive Data Synthesis For Universal Multimodal Retrieval}, 
      author={Junjie Zhou and Zheng Liu and Ze Liu and Shitao Xiao and Yueze Wang and Bo Zhao and Chen Jason Zhang and Defu Lian and Yongping Xiong},
      year={2024},
      eprint={2412.14475},
      archivePrefix={arXiv},
      primaryClass={cs.CV},
      url={https://arxiv.org/abs/2412.14475}, 
}

@inproceedings{ma2024dse,
    title = "Unifying Multimodal Retrieval via Document Screenshot Embedding",
    author = "Ma, Xueguang  and
      Lin, Sheng-Chieh  and
      Li, Minghan  and
      Chen, Wenhu  and
      Lin, Jimmy",
    editor = "Al-Onaizan, Yaser  and
      Bansal, Mohit  and
      Chen, Yun-Nung",
    booktitle = "Proceedings of the 2024 Conference on Empirical Methods in Natural Language Processing",
    month = nov,
    year = "2024",
    address = "Miami, Florida, USA",
    publisher = "Association for Computational Linguistics",
    url = "https://aclanthology.org/2024.emnlp-main.373/",
    doi = "10.18653/v1/2024.emnlp-main.373",
    pages = "6492--6505",
}

@misc{lan2025llave,
      title={LLaVE: Large Language and Vision Embedding Models with Hardness-Weighted Contrastive Learning}, 
      author={Zhibin Lan and Liqiang Niu and Fandong Meng and Jie Zhou and Jinsong Su},
      year={2025},
      eprint={2503.04812},
      archivePrefix={arXiv},
      primaryClass={cs.CV},
      url={https://arxiv.org/abs/2503.04812}, 
}

@misc{liu2024lamra,
      title={LamRA: Large Multimodal Model as Your Advanced Retrieval Assistant}, 
      author={Yikun Liu and Pingan Chen and Jiayin Cai and Xiaolong Jiang and Yao Hu and Jiangchao Yao and Yanfeng Wang and Weidi Xie},
      year={2024},
      eprint={2412.01720},
      archivePrefix={arXiv},
      primaryClass={cs.CV},
      url={https://arxiv.org/abs/2412.01720}, 
}

@misc{zhang2025ocrhindersrag,
      title={OCR Hinders RAG: Evaluating the Cascading Impact of OCR on Retrieval-Augmented Generation}, 
      author={Junyuan Zhang and Qintong Zhang and Bin Wang and Linke Ouyang and Zichen Wen and Ying Li and Ka-Ho Chow and Conghui He and Wentao Zhang},
      year={2025},
      eprint={2412.02592},
      archivePrefix={arXiv},
      primaryClass={cs.CV},
      url={https://arxiv.org/abs/2412.02592}, 
}

@misc{chen2024tokenpredictionmultimodalintelligence,
      title={Next Token Prediction Towards Multimodal Intelligence: A Comprehensive Survey}, 
      author={Liang Chen and Zekun Wang and Shuhuai Ren and Lei Li and Haozhe Zhao and Yunshui Li and Zefan Cai and Hongcheng Guo and Lei Zhang and Yizhe Xiong and Yichi Zhang and Ruoyu Wu and Qingxiu Dong and Ge Zhang and Jian Yang and Lingwei Meng and Shujie Hu and Yulong Chen and Junyang Lin and Shuai Bai and Andreas Vlachos and Xu Tan and Minjia Zhang and Wen Xiao and Aaron Yee and Tianyu Liu and Baobao Chang},
      year={2024},
      eprint={2412.18619},
      archivePrefix={arXiv},
      primaryClass={cs.CL},
      url={https://arxiv.org/abs/2412.18619}, 
}

@InProceedings{radford2021clip,
  title = 	 {Learning Transferable Visual Models From Natural Language Supervision},
  author =       {Radford, Alec and Kim, Jong Wook and Hallacy, Chris and Ramesh, Aditya and Goh, Gabriel and Agarwal, Sandhini and Sastry, Girish and Askell, Amanda and Mishkin, Pamela and Clark, Jack and Krueger, Gretchen and Sutskever, Ilya},
  booktitle = 	 {Proceedings of the 38th International Conference on Machine Learning},
  pages = 	 {8748--8763},
  year = 	 {2021},
  editor = 	 {Meila, Marina and Zhang, Tong},
  volume = 	 {139},
  series = 	 {Proceedings of Machine Learning Research},
  month = 	 {18--24 Jul},
  publisher =    {PMLR},
  url = 	 {https://proceedings.mlr.press/v139/radford21a.html},
}

@InProceedings{li2022blip,
  title = 	 {{BLIP}: Bootstrapping Language-Image Pre-training for Unified Vision-Language Understanding and Generation},
  author =       {Li, Junnan and Li, Dongxu and Xiong, Caiming and Hoi, Steven},
  booktitle = 	 {Proceedings of the 39th International Conference on Machine Learning},
  pages = 	 {12888--12900},
  year = 	 {2022},
  editor = 	 {Chaudhuri, Kamalika and Jegelka, Stefanie and Song, Le and Szepesvari, Csaba and Niu, Gang and Sabato, Sivan},
  volume = 	 {162},
  series = 	 {Proceedings of Machine Learning Research},
  month = 	 {17--23 Jul},
  publisher =    {PMLR},
  url = 	 {https://proceedings.mlr.press/v162/li22n.html}
}

@InProceedings{li2023blip2,
  title = 	 {{BLIP}-2: Bootstrapping Language-Image Pre-training with Frozen Image Encoders and Large Language Models},
  author =       {Li, Junnan and Li, Dongxu and Savarese, Silvio and Hoi, Steven},
  booktitle = 	 {Proceedings of the 40th International Conference on Machine Learning},
  pages = 	 {19730--19742},
  year = 	 {2023},
  editor = 	 {Krause, Andreas and Brunskill, Emma and Cho, Kyunghyun and Engelhardt, Barbara and Sabato, Sivan and Scarlett, Jonathan},
  volume = 	 {202},
  series = 	 {Proceedings of Machine Learning Research},
  month = 	 {23--29 Jul},
  publisher =    {PMLR},
  url = 	 {https://proceedings.mlr.press/v202/li23q.html}
}

@InProceedings{zhai2023siglip,
    author    = {Zhai, Xiaohua and Mustafa, Basil and Kolesnikov, Alexander and Beyer, Lucas},
    title     = {Sigmoid Loss for Language Image Pre-Training},
    booktitle = {Proceedings of the IEEE/CVF International Conference on Computer Vision (ICCV)},
    month     = {October},
    year      = {2023},
    pages     = {11975-11986}
}

@InProceedings{cherti2023openclip,
    author    = {Cherti, Mehdi and Beaumont, Romain and Wightman, Ross and Wortsman, Mitchell and Ilharco, Gabriel and Gordon, Cade and Schuhmann, Christoph and Schmidt, Ludwig and Jitsev, Jenia},
    title     = {Reproducible Scaling Laws for Contrastive Language-Image Learning},
    booktitle = {Proceedings of the IEEE/CVF Conference on Computer Vision and Pattern Recognition (CVPR)},
    month     = {June},
    year      = {2023},
    pages     = {2818-2829}
}

@misc{zhang2024magiclens,
      title={MagicLens: Self-Supervised Image Retrieval with Open-Ended Instructions}, 
      author={Kai Zhang and Yi Luan and Hexiang Hu and Kenton Lee and Siyuan Qiao and Wenhu Chen and Yu Su and Ming-Wei Chang},
      year={2024},
      eprint={2403.19651},
      archivePrefix={arXiv},
      primaryClass={cs.CV},
      url={https://arxiv.org/abs/2403.19651}, 
}

@misc{riedler2024beyondtext,
      title={Beyond Text: Optimizing RAG with Multimodal Inputs for Industrial Applications}, 
      author={Monica Riedler and Stefan Langer},
      year={2024},
      eprint={2410.21943},
      archivePrefix={arXiv},
      primaryClass={cs.CL},
      url={https://arxiv.org/abs/2410.21943}, 
}

@misc{langchain2023multimodal,
  title     = {Multi-Modal RAG Template},
  author    = {{LangChain Team}},
  year      = {2023},
  month     = {Dec},
  day       = {6},
  url       = {https://blog.langchain.dev/multi-modal-rag-template/},
  publisher = {LangChain Blog}
}

@misc{nvidia2024multimodal,
  title     = {An Easy Introduction to Multimodal Retrieval-Augmented Generation},
  author    = {Annie Surla and Aditi Bodhankar and Tanay Varshney},
  year      = {2024},
  month     = {Mar},
  day       = {20},
  url       = {https://developer.nvidia.com/blog/an-easy-introduction-to-multimodal-retrieval-augmented-generation/},
  publisher = {NVIDIA Developer Blog}
}

@inproceedings{zhou2024vista,
    title = "{VISTA}: Visualized Text Embedding For Universal Multi-Modal Retrieval",
    author = "Zhou, Junjie  and
      Liu, Zheng  and
      Xiao, Shitao  and
      Zhao, Bo  and
      Xiong, Yongping",
    editor = "Ku, Lun-Wei  and
      Martins, Andre  and
      Srikumar, Vivek",
    booktitle = "Proceedings of the 62nd Annual Meeting of the Association for Computational Linguistics (Volume 1: Long Papers)",
    month = aug,
    year = "2024",
    address = "Bangkok, Thailand",
    publisher = "Association for Computational Linguistics",
    url = "https://aclanthology.org/2024.acl-long.175/",
    doi = "10.18653/v1/2024.acl-long.175",
    pages = "3185--3200"
}

@misc{oord2019infonce,
      title={Representation Learning with Contrastive Predictive Coding}, 
      author={Aaron van den Oord and Yazhe Li and Oriol Vinyals},
      year={2019},
      eprint={1807.03748},
      archivePrefix={arXiv},
      primaryClass={cs.LG},
      url={https://arxiv.org/abs/1807.03748}, 
}

@inproceedings{Khattab2020colbert,
    author = {Khattab, Omar and Zaharia, Matei},
    title = {ColBERT: Efficient and Effective Passage Search via Contextualized Late Interaction over BERT},
    year = {2020},
    isbn = {9781450380164},
    publisher = {Association for Computing Machinery},
    address = {New York, NY, USA},
    url = {https://doi.org/10.1145/3397271.3401075},
    doi = {10.1145/3397271.3401075},
    booktitle = {Proceedings of the 43rd International ACM SIGIR Conference on Research and Development in Information Retrieval},
    pages = {39–48},
    numpages = {10},
    location = {Virtual Event, China},
    series = {SIGIR '20}
}

@article{robertson2009bm25,
    author = {Robertson, Stephen and Zaragoza, Hugo},
    title = {The Probabilistic Relevance Framework: BM25 and Beyond},
    year = {2009},
    issue_date = {April 2009},
    publisher = {Now Publishers Inc.},
    address = {Hanover, MA, USA},
    volume = {3},
    number = {4},
    issn = {1554-0669},
    url = {https://doi.org/10.1561/1500000019},
    doi = {10.1561/1500000019},
    journal = {Found. Trends Inf. Retr.},
    month = apr,
    pages = {333–389},
    numpages = {57}
}

@misc{xiao2023bgeembedding,
      title={C-Pack: Packaged Resources To Advance General Chinese Embedding}, 
      author={Shitao Xiao and Zheng Liu and Peitian Zhang and Niklas Muennighoff},
      year={2023},
      eprint={2309.07597},
      archivePrefix={arXiv},
      primaryClass={cs.CL}
}

@inproceedings{
    hu2024minicpm,
    title={Mini{CPM}: Unveiling the Potential of Small Language Models with Scalable Training Strategies},
    author={Shengding Hu and Yuge Tu and Xu Han and Ganqu Cui and Chaoqun He and Weilin Zhao and Xiang Long and Zhi Zheng and Yewei Fang and Yuxiang Huang and Xinrong Zhang and Zhen Leng Thai and Chongyi Wang and Yuan Yao and Chenyang Zhao and Jie Zhou and Jie Cai and Zhongwu Zhai and Ning Ding and Chao Jia and Guoyang Zeng and dahai li and Zhiyuan Liu and Maosong Sun},
    booktitle={First Conference on Language Modeling},
    year={2024},
    url={https://openreview.net/forum?id=3X2L2TFr0f}
}

@inproceedings{
    cai2025matryoshka,
    title={Matryoshka Multimodal Models},
    author={Mu Cai and Jianwei Yang and Jianfeng Gao and Yong Jae Lee},
    booktitle={The Thirteenth International Conference on Learning Representations},
    year={2025},
    url={https://openreview.net/forum?id=Uhj5OxAz7I}
}

@misc{sturua2024jinaembeddingsv3,
      title={jina-embeddings-v3: Multilingual Embeddings With Task LoRA}, 
      author={Saba Sturua and Isabelle Mohr and Mohammad Kalim Akram and Michael Günther and Bo Wang and Markus Krimmel and Feng Wang and Georgios Mastrapas and Andreas Koukounas and Nan Wang and Han Xiao},
      year={2024},
      eprint={2409.10173},
      archivePrefix={arXiv},
      primaryClass={cs.CL},
      url={https://arxiv.org/abs/2409.10173}, 
}

@inproceedings{es-etal-2024-ragas,
    title = "{RAGA}s: Automated Evaluation of Retrieval Augmented Generation",
    author = "Es, Shahul  and
      James, Jithin  and
      Espinosa Anke, Luis  and
      Schockaert, Steven",
    editor = "Aletras, Nikolaos  and
      De Clercq, Orphee",
    booktitle = "Proceedings of the 18th Conference of the European Chapter of the Association for Computational Linguistics: System Demonstrations",
    month = mar,
    year = "2024",
    address = "St. Julians, Malta",
    publisher = "Association for Computational Linguistics",
    url = "https://aclanthology.org/2024.eacl-demo.16/",
    pages = "150--158",
}

@misc{saadfalcon2024ares,
      title={ARES: An Automated Evaluation Framework for Retrieval-Augmented Generation Systems}, 
      author={Jon Saad-Falcon and Omar Khattab and Christopher Potts and Matei Zaharia},
      year={2024},
      eprint={2311.09476},
      archivePrefix={arXiv},
      primaryClass={cs.CL},
      url={https://arxiv.org/abs/2311.09476}, 
}

@misc{mortaheb2025ragcheckevaluatingmultimodalretrieval,
      title={RAG-Check: Evaluating Multimodal Retrieval Augmented Generation Performance}, 
      author={Matin Mortaheb and Mohammad A. Amir Khojastepour and Srimat T. Chakradhar and Sennur Ulukus},
      year={2025},
      eprint={2501.03995},
      archivePrefix={arXiv},
      primaryClass={cs.LG},
      url={https://arxiv.org/abs/2501.03995}, 
}

@misc{yang2025benchmarkingmultimodalragchartbased,
      title={Benchmarking Multimodal RAG through a Chart-based Document Question-Answering Generation Framework}, 
      author={Yuming Yang and Jiang Zhong and Li Jin and Jingwang Huang and Jingpeng Gao and Qing Liu and Yang Bai and Jingyuan Zhang and Rui Jiang and Kaiwen Wei},
      year={2025},
      eprint={2502.14864},
      archivePrefix={arXiv},
      primaryClass={cs.AI},
      url={https://arxiv.org/abs/2502.14864}, 
}

@misc{liu2025benchmarkingretrievalaugmentedgenerationmultimodal,
      title={Benchmarking Retrieval-Augmented Generation in Multi-Modal Contexts}, 
      author={Zhenghao Liu and Xingsheng Zhu and Tianshuo Zhou and Xinyi Zhang and Xiaoyuan Yi and Yukun Yan and Yu Gu and Ge Yu and Maosong Sun},
      year={2025},
      eprint={2502.17297},
      archivePrefix={arXiv},
      primaryClass={cs.AI},
      url={https://arxiv.org/abs/2502.17297}, 
}

@inproceedings{abootorabi2025askinanymodality,
    title = "Ask in Any Modality: A Comprehensive Survey on Multimodal Retrieval-Augmented Generation",
    author = "Abootorabi, Mohammad Mahdi  and
      Zobeiri, Amirhosein  and
      Dehghani, Mahdi  and
      Mohammadkhani, Mohammadali  and
      Mohammadi, Bardia  and
      Ghahroodi, Omid  and
      Baghshah, Mahdieh Soleymani  and
      Asgari, Ehsaneddin",
    editor = "Che, Wanxiang  and
      Nabende, Joyce  and
      Shutova, Ekaterina  and
      Pilehvar, Mohammad Taher",
    booktitle = "Findings of the Association for Computational Linguistics: ACL 2025",
    month = jul,
    year = "2025",
    address = "Vienna, Austria",
    publisher = "Association for Computational Linguistics",
    url = "https://aclanthology.org/2025.findings-acl.861/",
    doi = "10.18653/v1/2025.findings-acl.861",
    pages = "16776--16809",
    ISBN = "979-8-89176-256-5",
}

@misc{mei2025mmragsurvey,
      title={A Survey of Multimodal Retrieval-Augmented Generation}, 
      author={Lang Mei and Siyu Mo and Zhihan Yang and Chong Chen},
      year={2025},
      eprint={2504.08748},
      archivePrefix={arXiv},
      primaryClass={cs.IR},
      url={https://arxiv.org/abs/2504.08748}, 
}

@misc{tamber2025benchmarkingllmfaithfulnessrag,
      title={Benchmarking LLM Faithfulness in RAG with Evolving Leaderboards}, 
      author={Manveer Singh Tamber and Forrest Sheng Bao and Chenyu Xu and Ge Luo and Suleman Kazi and Minseok Bae and Miaoran Li and Ofer Mendelevitch and Renyi Qu and Jimmy Lin},
      year={2025},
      eprint={2505.04847},
      archivePrefix={arXiv},
      primaryClass={cs.CL},
      url={https://arxiv.org/abs/2505.04847}, 
}

@InProceedings{vinyals2015showandtell,
    author = {Vinyals, Oriol and Toshev, Alexander and Bengio, Samy and Erhan, Dumitru},
    title = {Show and Tell: A Neural Image Caption Generator},
    booktitle = {Proceedings of the IEEE Conference on Computer Vision and Pattern Recognition (CVPR)},
    month = {June},
    year = {2015}
}

@InProceedings{xu2015showattendandtell,
  title = 	 {Show, Attend and Tell: Neural Image Caption Generation with Visual Attention},
  author = 	 {Xu, Kelvin and Ba, Jimmy and Kiros, Ryan and Cho, Kyunghyun and Courville, Aaron and Salakhudinov, Ruslan and Zemel, Rich and Bengio, Yoshua},
  booktitle = 	 {Proceedings of the 32nd International Conference on Machine Learning},
  pages = 	 {2048--2057},
  year = 	 {2015},
  editor = 	 {Bach, Francis and Blei, David},
  volume = 	 {37},
  series = 	 {Proceedings of Machine Learning Research},
  address = 	 {Lille, France},
  month = 	 {07--09 Jul},
  publisher =    {PMLR},
  url = 	 {https://proceedings.mlr.press/v37/xuc15.html},
}

@InProceedings{johnson2016densecap,
    author = {Johnson, Justin and Karpathy, Andrej and Fei-Fei, Li},
    title = {DenseCap: Fully Convolutional Localization Networks for Dense Captioning},
    booktitle = {Proceedings of the IEEE Conference on Computer Vision and Pattern Recognition (CVPR)},
    month = {June},
    year = {2016}
}

@misc{sidorov2020textcaps,
      title={TextCaps: a Dataset for Image Captioning with Reading Comprehension}, 
      author={Oleksii Sidorov and Ronghang Hu and Marcus Rohrbach and Amanpreet Singh},
      year={2020},
      eprint={2003.12462},
      archivePrefix={arXiv},
      primaryClass={cs.CV},
      url={https://arxiv.org/abs/2003.12462}, 
}

@inproceedings{xu2020layoutlm,
    author = {Xu, Yiheng and Li, Minghao and Cui, Lei and Huang, Shaohan and Wei, Furu and Zhou, Ming},
    title = {LayoutLM: Pre-training of Text and Layout for Document Image Understanding},
    year = {2020},
    isbn = {9781450379984},
    publisher = {Association for Computing Machinery},
    address = {New York, NY, USA},
    url = {https://doi.org/10.1145/3394486.3403172},
    doi = {10.1145/3394486.3403172},
    booktitle = {Proceedings of the 26th ACM SIGKDD International Conference on Knowledge Discovery \& Data Mining},
    pages = {1192–1200},
    numpages = {9},
    location = {Virtual Event, CA, USA},
    series = {KDD '20}
}

@inproceedings{xu2021layoutlmv2,
    title = "{L}ayout{LM}v2: Multi-modal Pre-training for Visually-rich Document Understanding",
    author = "Xu, Yang  and
      Xu, Yiheng  and
      Lv, Tengchao  and
      Cui, Lei  and
      Wei, Furu  and
      Wang, Guoxin  and
      Lu, Yijuan  and
      Florencio, Dinei  and
      Zhang, Cha  and
      Che, Wanxiang  and
      Zhang, Min  and
      Zhou, Lidong",
    editor = "Zong, Chengqing  and
      Xia, Fei  and
      Li, Wenjie  and
      Navigli, Roberto",
    booktitle = "Proceedings of the 59th Annual Meeting of the Association for Computational Linguistics and the 11th International Joint Conference on Natural Language Processing (Volume 1: Long Papers)",
    month = aug,
    year = "2021",
    address = "Online",
    publisher = "Association for Computational Linguistics",
    url = "https://aclanthology.org/2021.acl-long.201/",
    doi = "10.18653/v1/2021.acl-long.201",
    pages = "2579--2591",
}

@inproceedings{huang2022layoutlmv3,
    author = {Huang, Yupan and Lv, Tengchao and Cui, Lei and Lu, Yutong and Wei, Furu},
    title = {LayoutLMv3: Pre-training for Document AI with Unified Text and Image Masking},
    year = {2022},
    isbn = {9781450392037},
    publisher = {Association for Computing Machinery},
    address = {New York, NY, USA},
    url = {https://doi.org/10.1145/3503161.3548112},
    doi = {10.1145/3503161.3548112},
    booktitle = {Proceedings of the 30th ACM International Conference on Multimedia},
    pages = {4083–4091},
    numpages = {9},
    location = {Lisboa, Portugal},
    series = {MM '22}
}

@InProceedings{mathew2021docvqa,
    author    = {Mathew, Minesh and Karatzas, Dimosthenis and Jawahar, C.V.},
    title     = {DocVQA: A Dataset for VQA on Document Images},
    booktitle = {Proceedings of the IEEE/CVF Winter Conference on Applications of Computer Vision (WACV)},
    month     = {January},
    year      = {2021},
    pages     = {2200-2209}
}

@inproceedings{kim2022donut,
    author = {Kim, Geewook and Hong, Teakgyu and Yim, Moonbin and Nam, JeongYeon and Park, Jinyoung and Yim, Jinyeong and Hwang, Wonseok and Yun, Sangdoo and Han, Dongyoon and Park, Seunghyun},
    title = {OCR-Free Document Understanding Transformer},
    year = {2022},
    isbn = {978-3-031-19814-4},
    publisher = {Springer-Verlag},
    address = {Berlin, Heidelberg},
    url = {https://doi.org/10.1007/978-3-031-19815-1_29},
    doi = {10.1007/978-3-031-19815-1_29},
    booktitle = {Computer Vision – ECCV 2022: 17th European Conference, Tel Aviv, Israel, October 23–27, 2022, Proceedings, Part XXVIII},
    pages = {498–517},
    numpages = {20},
    location = {Tel Aviv, Israel}
}

@InProceedings{lee2022pix2struct,
  title = 	 {{P}ix2{S}truct: Screenshot Parsing as Pretraining for Visual Language Understanding},
  author =       {Lee, Kenton and Joshi, Mandar and Turc, Iulia Raluca and Hu, Hexiang and Liu, Fangyu and Eisenschlos, Julian Martin and Khandelwal, Urvashi and Shaw, Peter and Chang, Ming-Wei and Toutanova, Kristina},
  booktitle = 	 {Proceedings of the 40th International Conference on Machine Learning},
  pages = 	 {18893--18912},
  year = 	 {2023},
  editor = 	 {Krause, Andreas and Brunskill, Emma and Cho, Kyunghyun and Engelhardt, Barbara and Sabato, Sivan and Scarlett, Jonathan},
  volume = 	 {202},
  series = 	 {Proceedings of Machine Learning Research},
  month = 	 {23--29 Jul},
  publisher =    {PMLR},
  url = 	 {https://proceedings.mlr.press/v202/lee23g.html},
}

@InProceedings{miech2019howto100m,
    author = {Miech, Antoine and Zhukov, Dimitri and Alayrac, Jean-Baptiste and Tapaswi, Makarand and Laptev, Ivan and Sivic, Josef},
    title = {HowTo100M: Learning a Text-Video Embedding by Watching Hundred Million Narrated Video Clips},
    booktitle = {Proceedings of the IEEE/CVF International Conference on Computer Vision (ICCV)},
    month = {October},
    year = {2019}
}

@inproceedings{xu2021videoclip,
    title = "{V}ideo{CLIP}: Contrastive Pre-training for Zero-shot Video-Text Understanding",
    author = "Xu, Hu  and
      Ghosh, Gargi  and
      Huang, Po-Yao  and
      Okhonko, Dmytro  and
      Aghajanyan, Armen  and
      Metze, Florian  and
      Zettlemoyer, Luke  and
      Feichtenhofer, Christoph",
    editor = "Moens, Marie-Francine  and
      Huang, Xuanjing  and
      Specia, Lucia  and
      Yih, Scott Wen-tau",
    booktitle = "Proceedings of the 2021 Conference on Empirical Methods in Natural Language Processing",
    month = nov,
    year = "2021",
    address = "Online and Punta Cana, Dominican Republic",
    publisher = "Association for Computational Linguistics",
    url = "https://aclanthology.org/2021.emnlp-main.544/",
    doi = "10.18653/v1/2021.emnlp-main.544",
    pages = "6787--6800",
}

@inproceedings{lei2018tvqa,
    title = "{TVQA}: Localized, Compositional Video Question Answering",
    author = "Lei, Jie  and
      Yu, Licheng  and
      Bansal, Mohit  and
      Berg, Tamara",
    editor = "Riloff, Ellen  and
      Chiang, David  and
      Hockenmaier, Julia  and
      Tsujii, Jun{'}ichi",
    booktitle = "Proceedings of the 2018 Conference on Empirical Methods in Natural Language Processing",
    month = oct # "-" # nov,
    year = "2018",
    address = "Brussels, Belgium",
    publisher = "Association for Computational Linguistics",
    url = "https://aclanthology.org/D18-1167/",
    doi = "10.18653/v1/D18-1167",
    pages = "1369--1379",
}

@misc{li2018spokensquad,
      title={Spoken SQuAD: A Study of Mitigating the Impact of Speech Recognition Errors on Listening Comprehension}, 
      author={Chia-Hsuan Li and Szu-Lin Wu and Chi-Liang Liu and Hung-yi Lee},
      year={2018},
      eprint={1804.00320},
      archivePrefix={arXiv},
      primaryClass={cs.CL},
      url={https://arxiv.org/abs/1804.00320}, 
}

@inproceedings{kim2019audiocaps,
    title = "{A}udio{C}aps: Generating Captions for Audios in The Wild",
    author = "Kim, Chris Dongjoo  and
      Kim, Byeongchang  and
      Lee, Hyunmin  and
      Kim, Gunhee",
    editor = "Burstein, Jill  and
      Doran, Christy  and
      Solorio, Thamar",
    booktitle = "Proceedings of the 2019 Conference of the North {A}merican Chapter of the Association for Computational Linguistics: Human Language Technologies, Volume 1 (Long and Short Papers)",
    month = jun,
    year = "2019",
    address = "Minneapolis, Minnesota",
    publisher = "Association for Computational Linguistics",
    url = "https://aclanthology.org/N19-1011/",
    doi = "10.18653/v1/N19-1011",
    pages = "119--132",
}

@misc{drossos2019clotho,
      title={Clotho: An Audio Captioning Dataset}, 
      author={Konstantinos Drossos and Samuel Lipping and Tuomas Virtanen},
      year={2019},
      eprint={1910.09387},
      archivePrefix={arXiv},
      primaryClass={cs.SD},
      url={https://arxiv.org/abs/1910.09387}, 
}

@InProceedings{johnson2015imageretrievalusingscenegraphs,
    author = {Johnson, Justin and Krishna, Ranjay and Stark, Michael and Li, Li-Jia and Shamma, David and Bernstein, Michael and Fei-Fei, Li},
    title = {Image Retrieval Using Scene Graphs},
    booktitle = {Proceedings of the IEEE Conference on Computer Vision and Pattern Recognition (CVPR)},
    month = {June},
    year = {2015}
}

@InProceedings{chen2017mv3d,
    author = {Chen, Xiaozhi and Ma, Huimin and Wan, Ji and Li, Bo and Xia, Tian},
    title = {Multi-View 3D Object Detection Network for Autonomous Driving},
    booktitle = {Proceedings of the IEEE Conference on Computer Vision and Pattern Recognition (CVPR)},
    month = {July},
    year = {2017}
}

@InProceedings{Yang2018pixor,
    author = {Yang, Bin and Luo, Wenjie and Urtasun, Raquel},
    title = {PIXOR: Real-Time 3D Object Detection From Point Clouds},
    booktitle = {Proceedings of the IEEE Conference on Computer Vision and Pattern Recognition (CVPR)},
    month = {June},
    year = {2018}
}

@InProceedings{Lang2019PointPillars,
    author = {Lang, Alex H. and Vora, Sourabh and Caesar, Holger and Zhou, Lubing and Yang, Jiong and Beijbom, Oscar},
    title = {PointPillars: Fast Encoders for Object Detection From Point Clouds},
    booktitle = {Proceedings of the IEEE/CVF Conference on Computer Vision and Pattern Recognition (CVPR)},
    month = {June},
    year = {2019}
}

@inproceedings{chen2020r2gen,
    title = "Generating Radiology Reports via Memory-driven Transformer",
    author = "Chen, Zhihong  and
      Song, Yan  and
      Chang, Tsung-Hui  and
      Wan, Xiang",
    editor = "Webber, Bonnie  and
      Cohn, Trevor  and
      He, Yulan  and
      Liu, Yang",
    booktitle = "Proceedings of the 2020 Conference on Empirical Methods in Natural Language Processing (EMNLP)",
    month = nov,
    year = "2020",
    address = "Online",
    publisher = "Association for Computational Linguistics",
    url = "https://aclanthology.org/2020.emnlp-main.112/",
    doi = "10.18653/v1/2020.emnlp-main.112",
    pages = "1439--1449",
}

@misc{luo2024videorag,
      title={Video-RAG: Visually-aligned Retrieval-Augmented Long Video Comprehension}, 
      author={Yongdong Luo and Xiawu Zheng and Xiao Yang and Guilin Li and Haojia Lin and Jinfa Huang and Jiayi Ji and Fei Chao and Jiebo Luo and Rongrong Ji},
      year={2024},
      eprint={2411.13093},
      archivePrefix={arXiv},
      primaryClass={cs.CV},
      url={https://arxiv.org/abs/2411.13093}, 
}

@misc{zeng2025scenerag,
      title={SceneRAG: Scene-level Retrieval-Augmented Generation for Video Understanding}, 
      author={Nianbo Zeng and Haowen Hou and Fei Richard Yu and Si Shi and Ying Tiffany He},
      year={2025},
      eprint={2506.07600},
      archivePrefix={arXiv},
      primaryClass={cs.CV},
      url={https://arxiv.org/abs/2506.07600}, 
}

@inproceedings{fan2024videoagent,
    author = {Fan, Yue and Ma, Xiaojian and Wu, Rujie and Du, Yuntao and Li, Jiaqi and Gao, Zhi and Li, Qing},
    title = {VideoAgent: A Memory-Augmented Multimodal Agent for Video Understanding},
    year = {2024},
    isbn = {978-3-031-72669-9},
    publisher = {Springer-Verlag},
    address = {Berlin, Heidelberg},
    url = {https://doi.org/10.1007/978-3-031-72670-5_5},
    doi = {10.1007/978-3-031-72670-5_5},
    booktitle = {Computer Vision – ECCV 2024: 18th European Conference, Milan, Italy, September 29 – October 4, 2024, Proceedings,  Part XXII},
    pages = {75–92},
    numpages = {18},
    location = {Milan, Italy}
}

@inproceedings{jeong2025videorag,
    title = "{V}ideo{RAG}: Retrieval-Augmented Generation over Video Corpus",
    author = "Jeong, Soyeong  and
      Kim, Kangsan  and
      Baek, Jinheon  and
      Hwang, Sung Ju",
    editor = "Che, Wanxiang  and
      Nabende, Joyce  and
      Shutova, Ekaterina  and
      Pilehvar, Mohammad Taher",
    booktitle = "Findings of the Association for Computational Linguistics: ACL 2025",
    month = jul,
    year = "2025",
    address = "Vienna, Austria",
    publisher = "Association for Computational Linguistics",
    url = "https://aclanthology.org/2025.findings-acl.1096/",
    doi = "10.18653/v1/2025.findings-acl.1096",
    pages = "21278--21298",
    ISBN = "979-8-89176-256-5",
}

@misc{ren2025videorag,
      title={VideoRAG: Retrieval-Augmented Generation with Extreme Long-Context Videos}, 
      author={Xubin Ren and Lingrui Xu and Long Xia and Shuaiqiang Wang and Dawei Yin and Chao Huang},
      year={2025},
      eprint={2502.01549},
      archivePrefix={arXiv},
      primaryClass={cs.IR},
      url={https://arxiv.org/abs/2502.01549}, 
}

@InProceedings{Ventura2025chapterllama,
    author    = {Ventura, Lucas and Yang, Antoine and Schmid, Cordelia and Varol, G\"ul},
    title     = {Chapter-Llama: Efficient Chaptering in Hour-Long Videos with LLMs},
    booktitle = {Proceedings of the IEEE/CVF Conference on Computer Vision and Pattern Recognition (CVPR)},
    month     = {June},
    year      = {2025},
    pages     = {18947-18958}
}

@inproceedings{yang2023vidchapters7m,
    author = {Yang, Antoine and Nagrani, Arsha and Laptev, Ivan and Sivic, Josef and Schmid, Cordelia},
    booktitle = {Advances in Neural Information Processing Systems},
    editor = {A. Oh and T. Naumann and A. Globerson and K. Saenko and M. Hardt and S. Levine},
    pages = {49428--49444},
    publisher = {Curran Associates, Inc.},
    title = {VidChapters-7M: Video Chapters at Scale},
    url = {https://proceedings.neurips.cc/paper_files/paper/2023/file/9b5c3e00d6ed30aad7adac9e7a664de1-Paper-Datasets_and_Benchmarks.pdf},
    volume = {36},
    year = {2023}
}

@misc{zheng2025ragandunderstandinginvision,
      title={Retrieval Augmented Generation and Understanding in Vision: A Survey and New Outlook}, 
      author={Xu Zheng and Ziqiao Weng and Yuanhuiyi Lyu and Lutao Jiang and Haiwei Xue and Bin Ren and Danda Paudel and Nicu Sebe and Luc Van Gool and Xuming Hu},
      year={2025},
      eprint={2503.18016},
      archivePrefix={arXiv},
      primaryClass={cs.CV},
      url={https://arxiv.org/abs/2503.18016}, 
}

@InProceedings{jia2021ALIGN,
  title = 	 {Scaling Up Visual and Vision-Language Representation Learning With Noisy Text Supervision},
  author =       {Jia, Chao and Yang, Yinfei and Xia, Ye and Chen, Yi-Ting and Parekh, Zarana and Pham, Hieu and Le, Quoc and Sung, Yun-Hsuan and Li, Zhen and Duerig, Tom},
  booktitle = 	 {Proceedings of the 38th International Conference on Machine Learning},
  pages = 	 {4904--4916},
  year = 	 {2021},
  editor = 	 {Meila, Marina and Zhang, Tong},
  volume = 	 {139},
  series = 	 {Proceedings of Machine Learning Research},
  month = 	 {18--24 Jul},
  publisher =    {PMLR},
  url = 	 {https://proceedings.mlr.press/v139/jia21b.html},
}

@InProceedings{Zhai2022LiT,
    author    = {Zhai, Xiaohua and Wang, Xiao and Mustafa, Basil and Steiner, Andreas and Keysers, Daniel and Kolesnikov, Alexander and Beyer, Lucas},
    title     = {LiT: Zero-Shot Transfer With Locked-Image Text Tuning},
    booktitle = {Proceedings of the IEEE/CVF Conference on Computer Vision and Pattern Recognition (CVPR)},
    month     = {June},
    year      = {2022},
    pages     = {18123-18133}
}

@misc{yu2022coca,
      title={CoCa: Contrastive Captioners are Image-Text Foundation Models}, 
      author={Jiahui Yu and Zirui Wang and Vijay Vasudevan and Legg Yeung and Mojtaba Seyedhosseini and Yonghui Wu},
      year={2022},
      eprint={2205.01917},
      archivePrefix={arXiv},
      primaryClass={cs.CV},
      url={https://arxiv.org/abs/2205.01917}, 
}

@misc{yao2021filip,
      title={FILIP: Fine-grained Interactive Language-Image Pre-Training}, 
      author={Lewei Yao and Runhui Huang and Lu Hou and Guansong Lu and Minzhe Niu and Hang Xu and Xiaodan Liang and Zhenguo Li and Xin Jiang and Chunjing Xu},
      year={2021},
      eprint={2111.07783},
      archivePrefix={arXiv},
      primaryClass={cs.CV},
      url={https://arxiv.org/abs/2111.07783}, 
}

@InProceedings{andrew2013deepcca,
  title = 	 {Deep Canonical Correlation Analysis},
  author = 	 {Andrew, Galen and Arora, Raman and Bilmes, Jeff and Livescu, Karen},
  booktitle = 	 {Proceedings of the 30th International Conference on Machine Learning},
  pages = 	 {1247--1255},
  year = 	 {2013},
  editor = 	 {Dasgupta, Sanjoy and McAllester, David},
  volume = 	 {28},
  number =       {3},
  series = 	 {Proceedings of Machine Learning Research},
  address = 	 {Atlanta, Georgia, USA},
  month = 	 {17--19 Jun},
  publisher =    {PMLR},
  url = 	 {https://proceedings.mlr.press/v28/andrew13.html},
}

@InProceedings{Lee2018SCAN,
    author = {Lee, Kuang-Huei and Chen, Xi and Hua, Gang and Hu, Houdong and He, Xiaodong},
    title = {Stacked Cross Attention for Image-Text Matching},
    booktitle = {Proceedings of the European Conference on Computer Vision (ECCV)},
    month = {September},
    year = {2018}
}

\appendix

\section{Supplemental Data}
\label{sec:appendix-performance}

\begin{table}
  \centering
  \small
  \rowcolors{2}{gray!10}{white}
  \begin{tabular}{l c c c c c c}
    \toprule
    \multirow{2}{*}{\textbf{Model}} & \multicolumn{3}{c}{\textbf{Offline (ms)}} & \multicolumn{3}{c}{\textbf{Online (ms)}} \\ 
    \cmidrule(lr){2-4} \cmidrule(lr){5-7} & P. & E.  & Total & E. & S. & Total \\
    \midrule
    MiniCPM   & 284     & 28       & 312   & 28       & 26     & 54    \\
    VisRAG-Ret      & --      & 121      & 121   & 28       & 26     & 54    \\
    \bottomrule
  \end{tabular}
  \caption{Latency comparison between an OCR-reliant pipeline (MiniCPM \cite{hu2024minicpm}) and an MLLM-based end-to-end representer (VisRAG-Ret \cite{yu2025visrag}) during offline and online processing stages. Abbreviations: P. - Parsing; E. - Encoding; S. - Searching.}
  \label{tab:latency_comparison}
\end{table}

\begin{table*}[ht]
    \centering
    \resizebox{\textwidth}{!}{
        \begin{tabular}{lccccccc}
            \toprule
            \multirow{2}{*}{\textbf{Model}} & \multicolumn{4}{c}{\textbf{Per Meta-Task Score}} & & \multicolumn{1}{c}{\textbf{Average Score}} \\ 
            \cmidrule(lr){2-5} \cmidrule(lr){7-7}
                                   & Classification & VQA  & Retrieval & Grounding & & Overall \\ \midrule
            \# of datasets $\rightarrow$ & 10 & 10 & 12 & 4 & & 36 \\ \midrule
            
            \multicolumn{7}{c}{\emph{ Baselines }} \\
            \midrule
            \rowcolor{white}
            CLIP~\citep{radford2021clip}   & 42.8 & 9.1 &  53.0 &  51.8 &   &  37.8 \\
            \rowcolor{gray!10}
            BLIP2~\citep{li2023blip2}  & 27.0  &  4.2 & 33.9  & 47.0 &  &  25.2 \\
            \rowcolor{white}
            SigLIP~\citep{zhai2023siglip}  & 40.3  &  8.4 & 31.6  & 59.5 &  &  34.8 \\
            \rowcolor{gray!10}
            OpenCLIP~\citep{cherti2023openclip}  & \textbf{47.8}  &  10.9 & 52.3  & 53.3 &  &  39.7 \\
            \rowcolor{white}
            UniIR (BLIP\textsubscript{FF})~\citep{wei2024uniir} &  42.1 &	 \underline{15.0}  &	\underline{60.1} & 	\underline{62.2}  &	 & 	\underline{42.8} \\
            \rowcolor{gray!10}
            UniIR (CLIP\textsubscript{SF})~\citep{wei2024uniir} &  \underline{44.3} & \textbf{16.2} & \textbf{61.8} & \textbf{65.3} & &  \textbf{44.7}   \\
            \rowcolor{white}
            Magiclens~\citep{zhang2024magiclens}  &  38.8 &  8.3  &  35.4 &  26.0 &  & 27.8  \\
            \rowcolor{LightCyan}
            \textbf{\emph{Baseline Average}}   &  40.4 &  10.3  &  46.9 &  52.2  &  & 36.1  \\
            \midrule

            \multicolumn{7}{c}{\emph{ MLLMs as Multimodal Embedders }} \\ \midrule
            \rowcolor{white}
            VLM2Vec (Phi-3.5-V-4B) \cite{jiang2025vlm2vec}  & 54.8 & 54.9 & 62.3 & 79.5 & & 60.1  \\
            \rowcolor{gray!10}
            VLM2Vec (LLaVA-1.6-7B) \cite{jiang2025vlm2vec}  & 61.2 &	49.9	 & 67.4  &  \underline{86.1} & &  	62.9 \\
            \rowcolor{white}
            VLM2Vec (Qwen2-VL-7B) \cite{jiang2025vlm2vec}  & \underline{62.6} &	57.8	 & \underline{69.9}  &  81.7 & &  	\underline{65.8} \\
            \rowcolor{gray!10}
            MMRet-MLLM (LLaVA-1.6-7B) \cite{zhou2024mmret}  & 56.0 &	57.4	 & \underline{69.9}  &  83.6 & &  	64.1 \\
            \rowcolor{white}
            GME (Qwen2-VL-2B) \cite{zhang2025gme}  & 56.9 &	41.2	 & 67.8  &  53.4 & &  	55.8 \\
            \rowcolor{gray!10}
            LLaVE-2B \cite{lan2025llave}  & 62.1 &	\underline{60.2}	 & 65.2  &  84.9 & &  	65.2 \\
            \rowcolor{white}
            LLaVE-7B \cite{lan2025llave}  & \textbf{65.7} &	\textbf{65.4}	 & \textbf{70.9}  &  \textbf{91.9} & &  	\textbf{70.3} \\
            \rowcolor{LightCyan}
            \textbf{\emph{MLLM-based Average}}   &  59.9 &  55.3  &  67.6 &  80.2 &  & 63.5  \\
            \rowcolor{green!20}
            \textbf{\emph{Average Improvement ($\Delta$ = MLLM-based - Baselines) } }   &  +19.5 &  +45.0  &  +20.7 &  +28.0 &  & +27.4 \\
            \bottomrule
        \end{tabular}
    }
    \caption{ Performance comparison of multimodal embedding models on the MMEB benchmark, compiled from \cite{jiang2025vlm2vec} and other cited works. Scores are averaged per meta-task, and an overall average score is also provided. Within each model category, the best reported performance for each task is marked in \textbf{bold}, and the second-best is \underline{underlined}. This table synthesizes results to highlight the contrast between these model categories and summarizes the average improvement reported for MLLMs over the baselines. }
    \label{tab:mmeb_exp}
\end{table*}

\begin{table*}[ht]
    \centering
    \rowcolors{2}{gray!10}{white}
    \resizebox{\textwidth}{!}{
        \begin{tabular}{l c c c c c c c}
            \toprule
            \rowcolor{gray!20}
            \textbf{Model} & \textbf{ArxivQA} & \textbf{ChartQA} & \textbf{DocVQA} & \textbf{InfoVQA} & \textbf{PlotQA} & \textbf{SlideVQA} & \textbf{Average} \\
            \midrule
            
            \multicolumn{8}{c}{\textit{Baselines}} \\
            \midrule
            BM25 \citeyearpar{robertson2009bm25} (OCR)  & \underline{43.65} & 61.47 & \underline{75.27} & 66.94 & 57.28 & 86.78 & 65.23 \\
            bge-large \citeyearpar{xiao2023bgeembedding} (OCR) & 39.29 & 59.64 & 75.04 & 72.38 & 51.33 & 81.38 & 59.13 \\
            MiniCPM \citeyearpar{hu2024minicpm} (OCR) & 58.43 & 77.74 & 72.54 & \underline{83.45} & \textbf{64.78} & \underline{91.74} & \underline{74.78} \\
            NV-Embed-v2 \citeyearpar{lee2025nvembed} (OCR) & \textbf{59.39} & \underline{80.47} & \textbf{75.46} & \textbf{84.24} & 59.36 & \textbf{92.49} & \textbf{75.24} \\
            SigLIP \citeyearpar{zhai2023siglip} & \underline{59.16} & \textbf{81.34} & 64.60 & 74.59 & \underline{61.32} & 89.08 & 71.68 \\
            \midrule

            \multicolumn{8}{c}{\textit{MLLMs as End-to-End Representers}} \\
            \midrule
            ColPali \citeyearpar{faysse2025colpali} & \underline{72.50} & 73.49 & \textbf{82.79} & 81.15 & 55.32 & \textbf{93.99} & \underline{76.54} \\
            VisRAG-Ret \citeyearpar{yu2025visrag} & \textbf{75.11} & 76.63 & 75.37 & \textbf{86.37} & \underline{62.14} & 91.85 & \textbf{77.91} \\
            \bottomrule
        \end{tabular}
    }
    \caption{ Overall retrieval performance (MRR@10) across multiple Visual Question Answering (VQA) datasets, summarizing results from cited studies. This table synthesizes and compares reported performances of traditional baselines with \textit{MLLMs as End-to-End Representers}. In each model category, the best reported performance is marked in \textbf{bold}, and the second-best is \underline{underlined}. }
    \label{tab:retrieval_performance}
\end{table*}

\begin{figure}[t]
  \includegraphics[width=\columnwidth]{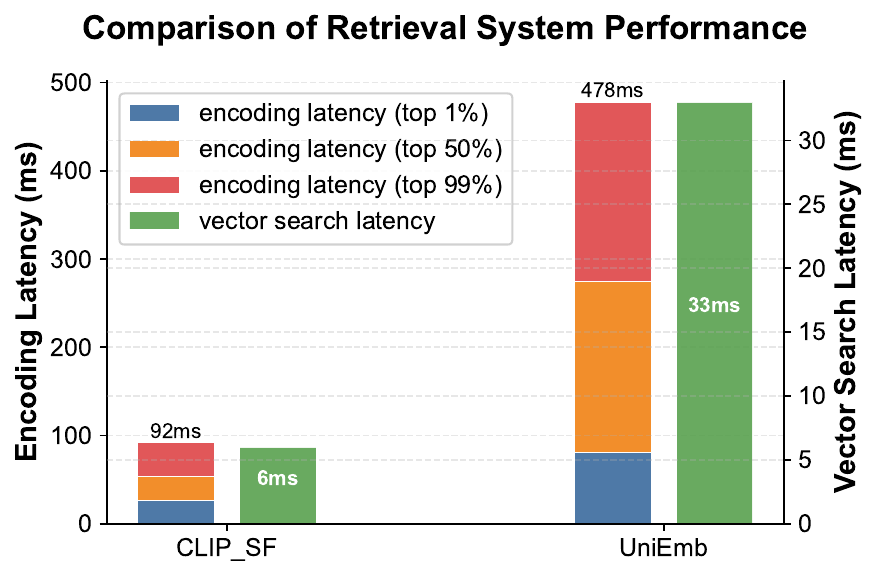}
  \caption{Comparison of encoding latency (displaying top 1\%, 50\%, and 99th percentiles) and vector search latency for the CLIP\textsubscript{SF} \cite{wei2024uniir} and UniEmb \cite{lin2025mmembed} models.
  Measurements were based on 100 randomly sampled queries from each of the 16 M-BEIR \cite{wei2024uniir} tasks.}
  \label{fig:retrieval-performance}
\end{figure}

This section provides supplementary empirical data referenced in the main survey, offering a more detailed view of the performance and cost trade-offs discussed.

\begin{itemize}[leftmargin=1.2em]
    \item \textbf{Table~\ref{tab:mmeb_exp}} presents a comprehensive comparison on the MMEB benchmark, substantiating the performance gains of the \emph{MLLM as Multimodal Embedder} role (\S\ref{subsec:mllm-as-mmemb}) over traditional baselines.
    
    \item \textbf{Table~\ref{tab:retrieval_performance}} details retrieval performance (MRR@10) across various VQA datasets, comparing \emph{End-to-End Representers} (\S\ref{subsec:mllm-as-e2erepr}) with baseline methods.
    
    \item \textbf{Table~\ref{tab:latency_comparison}} and \textbf{Figure~\ref{fig:retrieval-performance}} provide specific latency measurements, illustrating the computational overhead and costs discussed in \S\ref{subsec:computational_overhead_and_costs}.
\end{itemize}

\end{document}